%% file: paper.tex
\pdfoutput=1

\newif\iflocal
\localfalse

\documentclass[iop, apj, twocolappendix, numberedappendix]{emulateapj}
\usepackage{xspace}
\usepackage{amsmath}
\usepackage{framed} 
\usepackage{txfonts}
\usepackage{epstopdf}
\usepackage[usenames, dvipsnames]{color}
\usepackage{rotating}
\usepackage{natbib}
\usepackage{ulem}
\usepackage[colorlinks=true,urlcolor=blue,linkcolor=blue,citecolor=blue]{hyperref}
\usepackage{sistyle}
\SIthousandsep{,}

\iflocal
\input{../../../_LatexInclude/commands}

\input{../../../_LatexInclude/citation_fix}
\input{macros}
\def\figdir{figs}
\else

\input{commands}
\input{macros}
\input{citation_fix}
\def\figdir{.}
\fi

\shorttitle{\colossus}
\shortauthors{Diemer}

\journalinfo{The Astrophysical Journal Supplement Series, {\rm 239:35 (13pp), 2018 December}}
\submitted{Received 2018 July 31; revised 2018 October 10; accepted 2018 November 1; published 2018 December 18}

\begin{document}


\defcitealias{planck_14}{Planck Coll. 2014}
\defcitealias{planck_16}{Planck Coll. 2016}
\defcitealias{planck_18}{Planck Coll. 2018}


\title{Colossus: A python toolkit for cosmology, large-scale structure, and dark matter halos}
\author{Benedikt Diemer}

\affil{
Institute for Theory and Computation, Harvard-Smithsonian Center for Astrophysics, 60 Garden St., Cambridge, MA 02138, USA; benedikt.diemer@cfa.harvard.edu
}


\begin{abstract}
This paper introduces \colossus, a public, open-source python package for calculations related to cosmology, the large-scale structure (LSS) of matter in the universe, and the properties of dark matter halos. The code is designed to be fast and easy to use, with a coherent, well-documented user interface. The cosmology module implements Friedman–-Lemaitre–-Robertson–-Walker cosmologies including curvature, relativistic species, and different dark energy equations of state, and provides fast computations of the linear matter power spectrum, variance, and correlation function. The LSS module is concerned with the properties of peaks in Gaussian random fields and halos in a statistical sense, including their peak height, peak curvature, halo bias, and mass function. The halo module deals with spherical overdensity radii and masses, density profiles, concentration, and the splashback radius. To facilitate the rapid exploration of these quantities, \colossus implements more than $40$ different fitting functions from the literature. I discuss the core routines in detail, with particular emphasis on their accuracy. \colossus is available at \href{https://bitbucket.org/bdiemer/colossus}{bitbucket.org/bdiemer/colossus}.
\end{abstract}



\section{Introduction}
\label{sec:intro}

Over the past decade, python has emerged as the most popular programming language for data analysis in astronomy, particularly for computations that do not demand large amounts of computing power. Many of those calculations are non-trivial but need to be implemented time and time again. For example, in the commonly accepted $\Lambda$CDM cosmology, distances and times must be integrated numerically, increasing the risk of programming errors and numerical inaccuracies. As a remedy, a number of cosmology calculators have been presented \citep[e.g.,][]{wright_06, code_astropy_13, code_astropy_18}.

The sub-field of structure formation commonly relies on quantities that are even harder to compute accurately, such as, for example, the linear power spectrum of fluctuations, the variance of the density field, or the matter--matter correlation function. The corresponding integrals tend to converge slowly, making their evaluation inefficient in an interpreted language like python. When considering the structure of cold dark matter halos on smaller scales, other calculations occur frequently, such as, for example, the conversion between halo mass definitions, halo density profiles, and their concentrations. 

In this paper, I introduce \colossus, a python package that standardizes these calculations into a coherent, well-documented user interface (the name is an acronym for COsmology, haLO, and large-Scale StrUcture toolS). \colossus is not intended to be an all-encompassing library for structure formation but to provide a simple interface for basic calculations. For example, \colossus does not replicate the functionality of galaxy-halo modeling codes such as \textsc{HaloTools} \citep{hearin_17} or data analysis tools such as \textsc{NBodyKit} \citep{hand_17}, but it does compute many of the basic quantities that such codes rely on. In this vein, \colossus has been developed with the following design goals in mind.
\begin{itemize}
\item {\it Intuitive usage:} the interface should be as clear and simple as possible, allowing the user to evaluate complex quantities in one or a few lines of code. For this purpose, numerous fitting functions have been implemented.
\item {\it Performance:} computationally intensive routines are, wherever possible, approximated using smart interpolation tables that rely on as few data points as possible given a desired accuracy. These tables are stored on disk between executions.
\item {\it Stand-alone:} \colossus has no dependencies except for the standard numpy and scipy packages.
\item {\it Pure python:} \colossus does not contain any non-python code or any code that needs to be compiled, and can thus be installed by simply cloning the repository or with an installer such as pip. 
\item {\it Numpy compatibility:} virtually all \colossus functions accept both numbers and numpy arrays as input, and return results in the corresponding dimensions.
\item {\it Large range of validity:} \colossus tries to cover as wide a range of input parameters as possible. For example, the cosmology module works between redshifts of $-0.995$ and $200$.
\item {\it Consistent units:} \colossus follows a coherent set of physical units that are used for the input to and output from all functions.
\item {\it Reproducibility:} \colossus contains a suite of about $90$ unit tests that ensure that the code functions as expected on the user's machine and python distribution.
\end{itemize}
Notably, accuracy is not listed among these principles. Although \colossus naturally strives to be as accurate as possible, some functions trade accuracy for speed. Throughout the paper, the accuracy of functions and interpolation tables is listed carefully. 

\begin{deluxetable*}{lcccccclll}
\tablecaption{Pre-set cosmologies
\label{table:cosmo}}
\tablewidth{0pt}
\tablehead{
\colhead{ID} &
\colhead{rel} & 
\colhead{$H_0$} & 
\colhead{$\Omega_{\rm m}$} & 
\colhead{$\Omega_{\rm b}$} & 
\colhead{$n_{\rm s}$} & 
\colhead{$\sigma_8$} & 
\colhead{Reference} & 
\colhead{Comment}
}
\startdata
planck18 & yes & 67.66 & 0.3111 & 0.0490 & 0.9665 & 0.8102 & \citetalias{planck_18}, Table 2 & Best fit, with BAO (column 6) \\
planck18-only & yes & 67.36 & 0.3153 & 0.0493 & 0.9649 & 0.8111 & \citetalias{planck_18}, Table 2 & Best fit, Planck only (column 5) \\
planck15 & yes & 67.74 & 0.3089 & 0.0486 & 0.9667 & 0.8159 & \citetalias{planck_16}, Table 4 & Best fit, with external data (column 6) \\
planck15-only & yes & 67.81 & 0.3080 & 0.0484 & 0.9677 & 0.8149 & \citetalias{planck_16}, Table 4 & Best fit, Planck only (column 2) \\
planck13 & yes & 67.77 & 0.3071 & 0.0483 & 0.9611 & 0.8288 & \citetalias{planck_14}, Table 5 & Best fit, with external data \\
planck13-only & yes & 67.11 & 0.3175 & 0.0490 & 0.9624 & 0.8344 & \citetalias{planck_14}, Table 2 & Best fit, Planck only \\
WMAP9 & yes & 69.32 & 0.2865 & 0.0463 & 0.9608 & 0.8200 & \citealt{hinshaw_13},   Table 4 & Best fit, combined data \\
WMAP9-only & yes & 69.70 & 0.2814 & 0.0464 & 0.9710 & 0.8200 & \citealt{hinshaw_13},   Table 2 & Max. likelihood, WMAP only \\
WMAP9-ML & yes & 69.70 & 0.2821 & 0.0461 & 0.9646 & 0.8170 & \citealt{hinshaw_13},   Table 2 & Max. likelihood, combined data \\
WMAP7 & yes & 70.20 & 0.2743 & 0.0458 & 0.9680 & 0.8160 & \citealt{komatsu_11},   Table 1 & Best fit, with BAO and $H_0$ \\
WMAP7-only & yes & 70.30 & 0.2711 & 0.0451 & 0.9660 & 0.8090 & \citealt{komatsu_11},   Table 1 & Max. likelihood, WMAP only \\
WMAP7-ML & yes & 70.40 & 0.2715 & 0.0455 & 0.9670 & 0.8100 & \citealt{komatsu_11},   Table 1 & Max. likelihood, with BAO and $H_0$ \\
WMAP5 & yes & 70.50 & 0.2732 & 0.0456 & 0.9600 & 0.8120 & \citealt{komatsu_09},   Table 1 & Best fit, with BAO and SNe \\
WMAP5-only & yes & 72.40 & 0.2495 & 0.0432 & 0.9610 & 0.7870 & \citealt{komatsu_09},   Table 1 & Max. likelihood, WMAP only \\
WMAP5-ML & yes & 70.20 & 0.2769 & 0.0459 & 0.9620 & 0.8170 & \citealt{komatsu_09},   Table 1 & Max. likelihood, with BAO and SNe \\
WMAP3 & yes & 73.50 & 0.2342 & 0.0413 & 0.9510 & 0.7420 & \citealt{spergel_07},   Table 5 & Best fit, WMAP only \\
WMAP3-ML & yes & 73.20 & 0.2370 & 0.0414 & 0.9540 & 0.7560 & \citealt{spergel_07},   Table 2 & Max. likelihood, WMAP only \\
WMAP1 & yes & 72.00 & 0.2700 & 0.0463 & 0.9900 & 0.9000 & \citealt{spergel_03},   Table 7/4 & Best fit, WMAP only \\
WMAP1-ML & yes & 68.00 & 0.3136 & 0.0497 & 0.9700 & 0.9000 & \citealt{spergel_03},   Table 1/4 & Max. likelihood, WMAP only \\
illustris & no & 70.40 & 0.2726 & 0.0456 & 0.9630 & 0.8090 & \citealt{vogelsberger_14} & Cosmology of the Illustris simulation \\
bolshoi & no & 70.00 & 0.2700 & 0.0469 & 0.9500 & 0.8200 & \citealt{klypin_11} & Cosmology of the Bolshoi simulation \\
multidark-planck & no & 67.80 & 0.3070 & 0.0480 & 0.9600 & 0.8290 & \citealt{klypin_16} & Cosmology of the Multidark-Planck simulations \\
millennium & no & 73.00 & 0.2500 & 0.0450 & 1.0000 & 0.9000 & \citealt{springel_05_millennium} & Cosmology of the Millennium simulation \\
EdS & no & 70.00 & 1.0000 & 0.0000 & 1.0000 & 0.8200 & --- & Einstein-de Sitter cosmology
\enddata
\tablecomments{All cosmologies listed are flat $\Lambda$CDM cosmologies, i.e., they have no curvature and $w = -1$ (or no dark energy in the case of an Einstein-de Sitter cosmology). The ``rel'' field indicates whether relativistic species (neutrinos and radiation) are included. By default, they are included, except for the cosmologies of numerical simulations that do not explicitly follow relativistic species. The cosmology of the IllustrisTNG simulation suite is equivalent to the ``planck15'' cosmology \citep{pillepich_18}.}
\end{deluxetable*}

Sections \ref{sec:cosmo}--\ref{sec:halo} describe the main modules of \colossus in detail, Section~\ref{sec:conclusions} discusses future developments. The code repository is hosted at \href{https://bitbucket.org/bdiemer/colossus}{bitbucket.org/bdiemer/colossus}, but \colossus can also be automatically installed as a package by executing \texttt{pip install colossus}. An extensive online documentation is available at \href{https://bdiemer.bitbucket.io/colossus/}{bdiemer.bitbucket.io/colossus}. The code repository includes tutorials that explain how to use each module (in the form of Jupyter notebooks). This paper refers to code version 1.2.4.


\section{The Cosmology Module}
\label{sec:cosmo}

\begin{deluxetable*}{llcccccc}
\tablecaption{Accuracy and range of validity of key functions in the cosmology module
\label{table:cosmo_funcs}}
\tablewidth{0pt}
\tablehead{
\colhead{Function} &
\colhead{Dependent Funcs. (Incomplete List)} & 
\colhead{\S} & 
\colhead{Range $^a$} & 
\colhead{Comput. acc. $^b$} & 
\colhead{Interp. acc. $^b$} & 
\colhead{Inverse} & 
\colhead{Derivs. $^c$}
}
\startdata
Hubble parameter $E(z)$ & Densities, times & \ref{sec:cosmo:ez} & $-0.995 < z < 200$ & Mach. precision & - & no & - \\
Age of the universe & - & \ref{sec:cosmo:ez} & $-0.995 < z < 200$ & $10^{-8}$ & $6 \times 10^{-5}$ & yes & 1st, 2nd \\
Comoving distance & Angular diameter/luminosity dist. & \ref{sec:cosmo:ez} & $-0.995 < z < 200$ & $10^{-8}$ & - & yes & 1st, 2nd \\
Angular diameter dist. & - & \ref{sec:cosmo:ez} & $0 < z < 200$ & $10^{-8}$ & $9 \times 10^{-5}$ & no & 1st, 2nd \\
Luminosity distance & - & \ref{sec:cosmo:ez} & $0 < z < 200$ & $10^{-8}$ & $6 \times 10^{-4}$ & yes & 1st, 2nd \\
Linear growth factor & Power spectrum, variance, etc & \ref{sec:cosmo:dplus} & $-0.995 < z < 200$ & $10^{-6}$ (low $z$) $^d$ & $2 \times 10^{-4}$ & yes & 1st, 2nd \\
Linear power spectrum & Variance, correlation function & \ref{sec:cosmo:pk} & $10^{-20} < k < 10^{20}$ & $5\%$ (EH98) & $3 \times 10^{-4}$ & yes & 1st \\
Variance & Peak height, nonlinear mass & \ref{sec:cosmo:sigma} & $10^{-12} < R < 10^3$ & $3 \times 10^{-3}$ $^e$ & $5 \times 10^{-3}$ & yes & 1st \\
Correlation function & 2-halo term & \ref{sec:cosmo:xi} & $10^{-3} < R < 5 \times 10^2$ & $10^{-5}$ $^e$ & $10^{-2}$ & no & 1st
\enddata
\tablecomments{a) Many of the functions listed can be evaluated outside the given redshift range but have not been tested in those regimes. The ranges are given in the native units of the cosmology module, i.e., ${\rm Mpc}/h$ for distances and radii and $h/{\rm Mpc}$ for wavenumbers. A number of functions (the power spectrum, variance, and correlation function) depend on redshift through the linear growth factor, which thus defines the redshift range over which they can be reliably computed. \\
b) The table gives two different indicators of accuracy, namely the maximum error to which a quantity is computed (for example, in an integration), and the additional error due to interpolation. The latter can be avoided by switching interpolation off, though at a (sometimes steep) performance penalty. The interpolation error quoted here is the maximum error found at any redshift in a number of representative cosmologies. \\
c) Some functions can return their first derivative, and some functions return even higher orders. Note, however, that the accuracy of those derivatives is not guaranteed and likely to get worse at higher orders due to interpolation errors. In particular, cubic splines are prone to ringing that has virtually no effect on the solution but affects the higher derivatives of the spline.\\
d) The linear growth factor is computed to an integration accuracy of $10^{-6}$ at low redshift, but at high $z$ relies on the fitting function of \citet{gnedin_11_dc} for which no estimate of the accuracy is given. \\
e) The accuracy of the variance and correlation function refers to the accuracy of the integration and does not include the (generally much larger) inaccuracies due to the underlying approximation of the power spectrum (see Sections~\ref{sec:cosmo:sigma} and \ref{sec:cosmo:xi} for estimates of those errors).}
\end{deluxetable*}

The \colossus cosmology module implements the standard Friedman–-Lemaitre–-Robertson–-Walker cosmology and includes the contributions from dark matter, baryons, dark energy, curvature, photons, and neutrinos. The underlying expressions can be found in a number of cosmology textbooks \citep[e.g.,][]{rich_01, dodelson_03, ryden_03}.

\subsection{General Design}
\label{sec:cosmo:general}

In \colossus, the cosmology is set globally so that it does not need to be passed to functions. If necessary, the user can store multiple cosmology objects and activate them in turn. For convenience, the user can choose from an extensive list of pre-defined sets of cosmological parameters (Table~\ref{table:cosmo}). 

For many of its functions, the cosmology module relies on the interpolation of lookup tables to speed up the evaluation. These tables are computed on demand, i.e., when a function is first evaluated for a given cosmology. The number of bins in the tables was adjusted to obtain a particular accuracy. The interpolation uses cubic splines that are also used to invert the functions (e.g., give $z(t)$ rather than $t(z)$) and evaluate derivatives (e.g. $dz/dt$ or $dt/dz$). However, no guarantee is given on the accuracy of the derivatives, particularly for higher-order derivatives.

Table~\ref{table:cosmo_funcs} gives an overview of the accuracy and range of validity of key functions in the cosmology module. The quoted accuracy refers to the default parameters, but can be modified by the user in a number of ways. First, interpolation can be turned off altogether, though at a significant performance penalty. Second, the accuracy of many computations (such as numerical integrations) can be altered from the default settings. Finally, the user can change the binning scheme of the lookup tables, though this is generally not recommended.

Constructing {\it all} cosmological interpolation tables takes about $1.7$ s on a typical machine (2015 MacBook Pro, Python 3.5). This time is dominated by the correlation function without which the calculation is reduced to about $0.15$ s. Importantly, interpolation tables are computed only once: after the first execution, they are stored on disk and loaded on demand. To ensure that only matching tables are loaded for a given cosmology, all relevant parameters are expressed as a string and converted to a unique hash identifier using the md5 algorithm. The tables are discarded when cosmological parameters are changed.

\subsection{Initializing Cosmologies}
\label{sec:cosmo:init}

Cosmology objects are initiated from a set of cosmological parameters and settings, including the Hubble constant $H_0$ (often denoted $h \equiv H_0 / (100 {\rm km}/s/{\rm Mpc})$), the primordial power spectrum index $n_{\rm s}$, the power spectrum normalization $\sigma_8$, and the densities of certain species. The \colossus cosmology includes the densities of matter (dark matter and baryons), baryons, dark energy, curvature, photons, neutrinos, and the sum of relativistic species (photos and neutrinos), denoted $\rho_{\rm m}$, $\rho_{\rm b}$, $\rho_{\rm de}$, $\rho_{\rm k}$, $\rho_{\rm \gamma}$, $\rho_{\rm \nu}$, and $\rho_{\rm rel}$, respectively. Their fraction with respect to the critical density, $\rho_{\rm c}$, are denoted $\Omega_{\rm m}$, $\Omega_{\rm b}$, $\Omega_{\rm de}$, $\Omega_{\rm k}$, $\Omega_{\rm \gamma}$, $\Omega_{\rm \nu}$, and $\Omega_{\rm rel}$, their values at $z = 0$ as $\rho_{\rm m,0}$ and $\Omega_{\rm m,0}$ and so on. The user can specify whether a flat cosmology is assumed and whether relativistic species should be included. By default, the cosmology does contain relativistic species, namely radiation and neutrinos whose contributions are computed as follows. The density of radiation today follows from the Stefan--Boltzmann law for a blackbody: 
\begin{equation}
\rho_{\gamma,0} = 4 \sigma_{\rm SB}T_{\rm CMB,0}^4 \,,
\end{equation}
where $\sigma_{\rm SB}$ is the Stefan--Boltzmann constant. Dividing by the critical density and converting to the appropriate units, we find
\begin{equation}
\Omega_{\gamma,0} = \frac{4.4813 \times 10^{-7}}{h^2} \left( \frac{T_{\rm CMB,0}}{K} \right)^4 \approx 5.4 \times 10^{-5} \,,
\end{equation}
where we have assumed a default CMB temperature of $T_{\rm CMB,0} = 2.7255 K$ \citep{fixsen_09} and the ``planck15'' cosmology (Table~\ref{table:cosmo}). The density of neutrinos is
\begin{equation}
\Omega_{\nu,0} = \frac{7}{8} \left( \frac{4}{11} \right)^{4/3} N_{\rm eff} \Omega_{\gamma,0} \approx 0.69\, \Omega_{\gamma,0} \approx 3.7 \times 10^{-5} \,,
\end{equation}
where the effective number of neutrino species, $N_{\rm eff} = 3.046$, accounts for three neutrino species and subtle effects that influence the ratio of the photon and neutrino temperatures \citep{mangano_02, desalas_16, planck_18}. The user can, of course, set different values for $T_{\rm CMB,0}$ and $N_{\rm eff}$, but the neutrino mass is currently fixed. 

Finally, we need to compute the contributions from curvature and dark energy. If the cosmology is flat, $\Omega_{\rm k,0} = 0$ and 
\begin{equation}
\Omega_{\rm de,0} = 1 - \Omega_{\rm m,0} - \Omega_{\gamma,0} - \Omega_{\nu,0} \,.
\end{equation}
If the user has chosen a non-flat cosmology, $\Omega_{\rm de,0}$ needs to be set and we compute 
\begin{equation}
\Omega_{\rm k,0} = 1 - \Omega_{\rm de,0}  - \Omega_{\rm m,0} - \Omega_{\gamma,0} - \Omega_{\nu,0} \,.
\end{equation}
Dark energy is described by an equation of state parameter $w(z) = P(z)/\rho(z)$ which can be $-1$ (a cosmological constant resulting in a $\Lambda$CDM cosmology), a constant other than $-1$ (wCDM), linearly varying with redshift according to \citet{chevallier_01},
\begin{equation}
w(z) = w_0 + w_{\rm a} \frac{z}{z + 1} \,,
\end{equation}
or follow an arbitrary function $w(z)$ set by the user.

\subsection{Densities, Distances, Times}
\label{sec:cosmo:ez}

\begin{figure*}
\centering
\includegraphics[trim = 4mm 3mm 5mm 1mm, clip, scale=0.6]{\figdir/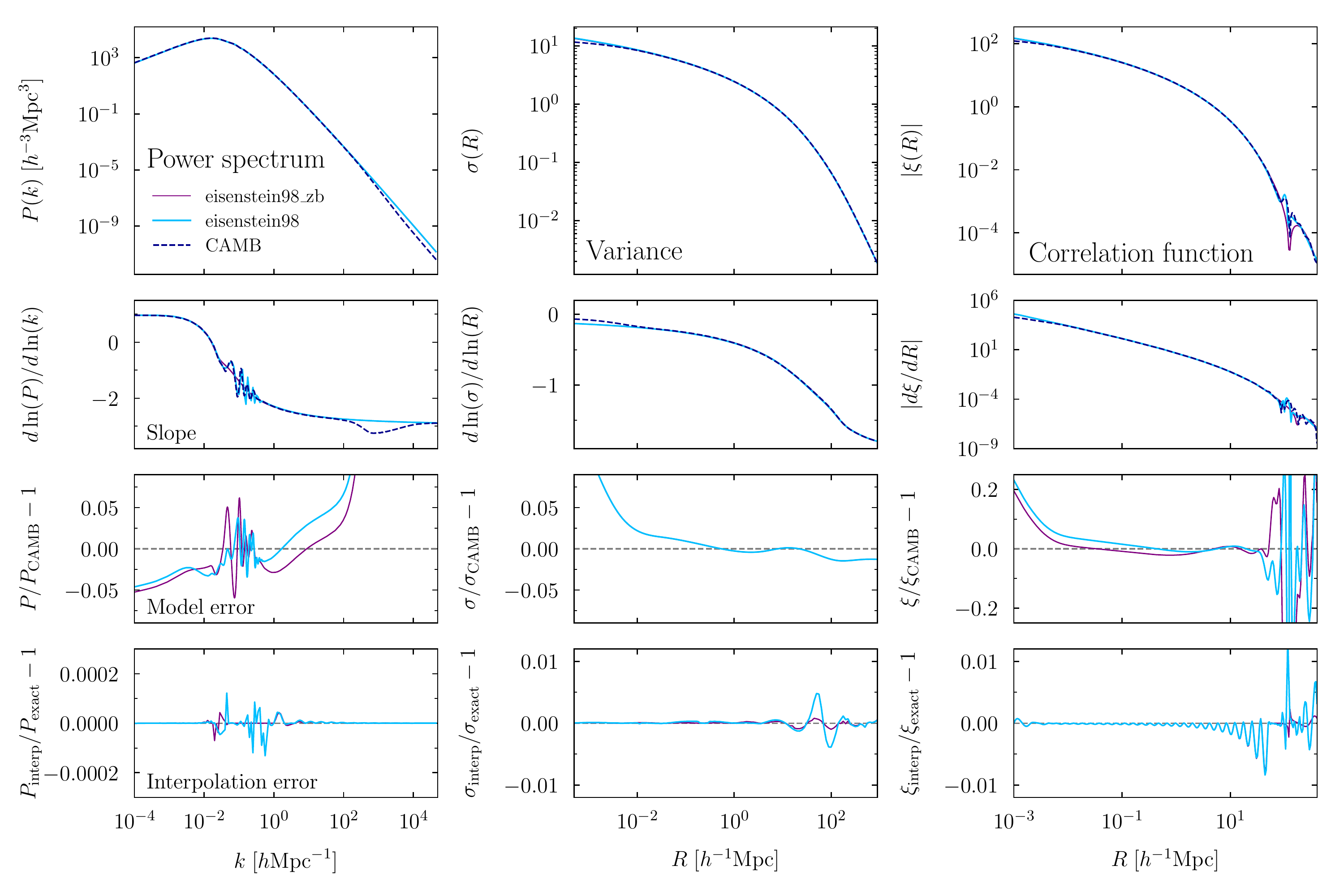}
\caption{The linear matter power spectrum, variance, and correlation function at $z = 0$ in the planck15 cosmology, computed numerically and from fitting functions. Left column: the top panel shows the power spectrum computed numerically using the \textsc{Camb} code \citep[][dashed dark blue line]{lewis_00} and the model of \citet[][light blue]{eisenstein_98} as well as their formula for the zero-baryon limit (purple). The zero-baryon case is not distinguishable by eye, but its slope is visibly different (second row), particularly around the wiggles due to the baryon acoustic oscillations (BAO). The \citet{eisenstein_98} function does not take the baryonic pressure into account, leading to excessive power at scales greater than $k \approx 100 h {\rm Mpc}^{-1}$. Elsewhere, the model reproduces the numerical calculation to better than 5\% (third row). This error contains a small interpolation error (shown in the bottom row as the ratio of the interpolated to the exact prediction of the \citet{eisenstein_98} model). This interpolation error is a function of the binning scheme used for the interpolation table and was designed to be insignificant. The $k$-range shown is the range over which the \textsc{Camb} calculation is defined, but the \colossus power spectrum can be evaluated at any $10^{-20} < k < 10^{20} h {\rm Mpc}^{-1}$. Center column: the variance, calculated using numerical integration (Equation~\ref{eq:sigma}). The \citet{eisenstein_98} approximation results in a variance that matches the numerical computation to better than 2\% except at very small radii where the additional power at small scales begins to matter. The numerically computed power spectrum would give increasingly poor results at small radii because of its limited $k$-range. The interpolation of $\sigma$ results in errors of at most $0.5\%$. Right column: same as center column but for the correlation function. The relative errors due to both the approximate power spectrum and interpolation grow around the zero-crossing, but the difference is small in absolute units.}
\label{fig:ps_sigma_xi}
\end{figure*}

The scale factor $a$ is defined as usual, $a = 1$ at $z = 0$. Many cosmological quantities rely on the Hubble constant as a function of time, normalized to the $z = 0$ value,
\begin{equation}
\label{eq:ez}
E(a) = \sqrt{\Omega_{\rm m,0} a^{-3} + \Omega_{\rm k,0} a^{-2} + \Omega_{\rm rel,0} a^{-4} + f(z) \Omega_{\rm de,0} } \,,
\end{equation}
where $f(z)$ encapsulates any possible evolution of the dark energy equation of state,
\begin{equation}
f(z) = \exp \left( 3 \int_0^{\ln(1+z)} [1 + w(z')] d \ln(1 + z') \right) \,.
\end{equation}
This expression evaluates to unity for a cosmological constant, to $f(z) = a^{-3 (1 + w_0)}$ for $w = w_0$, and to 
\begin{equation}
f(z) = a^{-3 (1 + w_0 + w_{\rm a})} e^{-3 w_{\rm a} (1 - a)}
\end{equation}
for $w(z) = w_0 + w_{\rm a} (1 - a)$ \citep{chevallier_01, linder_03_1}. \colossus always computes $E(z)$ exactly, i.e., without any interpolation. The critical density is evaluated as $\rho_{\rm c}(z) = \rho_{\rm c,0} E^2(z)$, whereas the other densities ($\rho_{\rm m}$, $\rho_{\rm b}$, $\rho_{\rm de}$, $\rho_{\gamma}$, $\rho_{\nu}$) are computed from their $z = 0$ values and the redshift scalings implied by Equation~(\ref{eq:ez}). A number of quantities depend on integrals of $E(z)$, and these quantities are stored in interpolation tables to avoid repeated evaluation of the integrals. For example, the age of the universe is
\begin{equation}
t(z) = \frac{1}{H_0} \int_z^{\infty} \frac{dz}{E(z) \times (1 + z)} \,.
\end{equation}
All times are expressed in units of Gyr in \colossus. The line-of-sight comoving distance to a particular redshift is
\begin{equation}
d_{\rm com,los}(z) = \frac{c}{H_0} \int_{0}^{z} \frac{1}{E(z)} \,,
\end{equation}
where all distances are expressed in units of comoving $\mpch$. The distance between two events at the same redshift that are separated by an angle of one radian depends on whether the cosmology is flat or not,
\begin{equation}
d_{\rm com,trans}(z) =  \left\{
\begin{array}{ll}
      \frac{c/H_0}{\sqrt{\Omega_{\rm k,0}}} \sinh \left(\frac{\sqrt{\Omega_{\rm k,0}}}{c/H_0} d_{\rm com,los} \right) & \forall \, \Omega_{\rm k,0} > 0 \\
      d_{\rm com,los} & \forall \, \Omega_{\rm k,0} = 0 \\
      \frac{c/H_0}{\sqrt{-\Omega_{\rm k,0}}} \sin \left(\frac{\sqrt{-\Omega_{\rm k,0}}}{c/H_0} d_{\rm com,los} \right) & \forall \, \Omega_{\rm k,0} < 0 \\
\end{array} 
\right. \,.
\end{equation}
In \colossus, this distance is referred to as the ``transverse comoving distance'' \citep[e.g.,][]{hogg_99}, but a number of other terms are used in the literature, e.g., ``comoving angular diameter distance'' \citep{dodelson_03}, ``comoving coordinate distance'' \citep{mo_10_book}, or ``angular size distance'' \citep{peebles_93}. The latter is not to be confused with the angular diameter distance $d_{\rm ang} = d_{\rm com,trans} / (1+z)$ or the luminosity distance $d_{\rm lum} = d_{\rm com,trans} (1+z)$.

The \colossus results for densities, distances, and times agree to a few times $10^{-4}$ or better with those computed by \astropy, which was itself tested against several other calculators \citep{code_astropy_13}. By default, interpolation is used to speed up the evaluations. For this purpose, \colossus computes tables with $50$ bins in redshift, equally spaced in $\ln(1 + z)$. The resulting interpolation errors are a few times $10^{-4}$ or better in all quantities, at all redshifts, and over a range of representative cosmologies (Table~\ref{table:cosmo_funcs}).

\subsection{The Linear Growth Factor}
\label{sec:cosmo:dplus}

\begin{deluxetable*}{lll}
\tablecaption{Fitting functions implemented in \colossus
\label{table:models}}
\tablewidth{0pt}
\tablehead{
\colhead{Model ID} &
\colhead{Reference} & 
\colhead{Comments}
}
\startdata
\multicolumn{3}{l}{\rule{0pt}{3ex} {\bf Cosmology: Power spectrum}} \\
\hline
\rule{0pt}{3ex} eisenstein98 & \citealt{eisenstein_98} & Semi-analytical fit to the transfer function calibrated based on numerical calculations \\
\rule{0pt}{0pt} eisenstein98\_zb & \citealt{eisenstein_98} & The zero-baryon limit of the \citealt{eisenstein_98} model (i.e., no baryon acoustic oscillations) \\
\multicolumn{3}{l}{\rule{0pt}{3ex} {\bf LSS: Mass function}} \\
\hline
\rule{0pt}{3ex} press74 & \citealt{press_74} & Prediction based on the statistics of peaks in Gaussian random fields (FOF, uses $\deltac(z)$) \\
\rule{0pt}{0pt} sheth99 & \citealt{sheth_99} & A calibration based on numerical results (FOF, uses $\deltac(z)$); see also \citealt{sheth_01} \\
\rule{0pt}{0pt} jenkins01 & \citealt{jenkins_01} & A calibration based on numerical results (FOF, no $z$-dependence) \\
\rule{0pt}{0pt} reed03 & \citealt{reed_03} & High-mass correction to \citealt{sheth_99} model (FOF, uses $\deltac(z)$) \\
\rule{0pt}{0pt} warren06 & \citealt{warren_06} & A calibration based on numerical results (FOF, no $z$-dependence) \\
\rule{0pt}{0pt} reed07 & \citealt{reed_07} & A model that takes the varying slope of the power spectrum into account (FOF, $z$-dependent) \\
\rule{0pt}{0pt} tinker08 & \citealt{tinker_08} & A calibration for SO halos with $200 \leq \Delta_{\rm m} \leq 3200$, explicit $z$-dependence \\
\rule{0pt}{0pt} crocce10 & \citealt{crocce_10} & A calibration based on numerical results (FOF, uses $\deltac(z)$) \\
\rule{0pt}{0pt} bhattacharya11 & \citealt{bhattacharya_11} & A calibration based on numerical results (FOF, explicit $z$-dependence) \\
\rule{0pt}{0pt} courtin11 & \citealt{courtin_11} & A calibration based on numerical results (FOF, uses fixed $\deltac = 1.673$) \\
\rule{0pt}{0pt} angulo12 & \citealt{angulo_12} & A calibration based on numerical results (FOF, no $z$-dependence) \\
\rule{0pt}{0pt} watson13 & \citealt{watson_13_mf} & Both FOF and SO fits, explicit $z$-dependence in the latter \\
\rule{0pt}{0pt} bocquet16 & \citealt{bocquet_16} & A model for different mass definitions, redshifts, and both hydro and DM-only simulations \\
\rule{0pt}{0pt} despali16 & \citealt{despali_16} & A redshift and mass definition-dependent calibration for both ellipsoidal and SO halo finders \\
\multicolumn{3}{l}{\rule{0pt}{3ex} {\bf LSS: Bias}} \\
\hline
\rule{0pt}{3ex} cole89 & \citealt{cole_89} & Bias prediction based on the peak-background split model \citep[see also][]{mo_96} \\
\rule{0pt}{0pt} jing98 & \citealt{jing_98} & Calibrated on scale-free universes, but also applicable to \LCDM \\
\rule{0pt}{0pt} sheth01 & \citealt{sheth_01} & Bias model taking the ellipsoidal nature of halos into account \\
\rule{0pt}{0pt} seljak04 & \citealt{seljak_04} & Model with an optional cosmological correction term \\
\rule{0pt}{0pt} pillepich10 & \citealt{pillepich_10} & A numerical calibration \\
\rule{0pt}{0pt} tinker10 & \citealt{tinker_10} & Fitting function that depends on the mass definition \\
\multicolumn{3}{l}{\rule{0pt}{3ex} {\bf Halo: Density profile}} \\
\hline
\rule{0pt}{3ex} Einasto & \citealt{einasto_65} & A three-parameter profile with smoothly varying slope \\
\rule{0pt}{0pt} Hernquist & \citealt{hernquist_90} & A two-parameter profile with inner and outer logarithmic slopes of $-1$ and $-4$ \\
\rule{0pt}{0pt} NFW & \citealt{navarro_95, navarro_96, navarro_97} & A two-parameter profile with inner and outer logarithmic slopes of $-1$ and $-3$ \\
\rule{0pt}{0pt} DK14 & \citealt{diemer_14} & A profile function that models the steepening due to the splashback radius \\
\multicolumn{3}{l}{\rule{0pt}{3ex} {\bf Halo: Concentration}} \\
\hline
\rule{0pt}{3ex} bullock01 & \citealt{bullock_01} & Universal model, $\ctoc$ for any mass, redshift, and cosmology \\
\rule{0pt}{0pt} duffy08 & \citealt{duffy_08} & Power-law fit, $\ctoc$, $\cvir$, and $\ctom$ for $1\times10^{11} < M < 10^{15} \msunh$, $0 < z < 2$, WMAP5 \\
\rule{0pt}{0pt} klypin11 & \citealt{klypin_11} & Power-law fit, $\cvir$ for $3\times10^{10} < M < 5\times10^{14} \msunh$, $z = 0$, WMAP7 cosmology \\
\rule{0pt}{0pt} prada12 & \citealt{prada_12} & Fit based on peak height, $\ctoc$ for any mass, redshift, and cosmology \\
\rule{0pt}{0pt} bhattacharya13 & \citealt{bhattacharya_13} & Power-law fit in $\nu$, $\ctoc$, $\cvir$, and $\ctom$ for $2\times10^{12} < M < 2\times10^{15} \msunh$, $0 < z < 2$, WMAP7 \\
\rule{0pt}{0pt} dutton14 & \citealt{dutton_14} & Power-law fit, $\ctoc$ and $\cvir$ for $M > 10^{10} \msunh$, $0 < z < 5$, planck13 cosmology \\
\rule{0pt}{0pt} diemer15 & \citealt{diemer_15} & Universal model, $\ctoc$ for any mass, redshift, or cosmology \\
\rule{0pt}{0pt} klypin16\_m & \citealt{klypin_16} & Power-law fit, $\ctoc$ and $\cvir$ for $M > 10^{10} \msunh$, $0 < z < 5$, planck13 or WMAP7 (function of $M$) \\
\rule{0pt}{0pt} klypin16\_nu & \citealt{klypin_16} & Power-law fit, $\ctoc$ and $\cvir$ for $M > 10^{10} \msunh$, $0 < z < 5$, planck13 (function of $\nu$) \\
\rule{0pt}{0pt} ludlow16 & \citealt{ludlow_16} & Universal model, $\ctoc$ for any mass, redshift, or cosmology \\
\rule{0pt}{0pt} child18 & \citealt{child_18} & Fit in $M/M_*$ space, $\ctoc$ for $M > 2.1 \times 10^{11} \msunh$, $0 < z < 4$, WMAP7 \\
\rule{0pt}{0pt} diemer18 & \citealt{diemer_18_cm} & Universal model, $\ctoc$ for any mass, redshift, or cosmology \\
\multicolumn{3}{l}{\rule{0pt}{3ex} {\bf Halo: Splashback radius}} \\
\hline
\rule{0pt}{3ex} adhikari14 & \citealt{adhikari_14} & Semi-analytical model, $\rsp$ and $\msp$ as a function of ($\Gamma$, $z$) \\
\rule{0pt}{0pt} more15 & \citealt{more_15} & Numerical calibration, $\rsp$ and $\msp$ as a function of ($\Gamma$, $z$) or ($M$, $z$) \\
\rule{0pt}{0pt} shi16 & \citealt{shi_16} & Semi-analytical model, $\rsp$ and $\msp$ as a function of ($\Gamma$, $z$) \\
\rule{0pt}{0pt} mansfield17 & \citealt{mansfield_17} & Numerical calibration, $\rsp$, $\msp$, and scatter as a function of ($\Gamma$, $M$, $z$) \\
\rule{0pt}{0pt} diemer17 & \citealt{diemer_17_rsp} & Numerical calibration, $\rsp$, $\msp$, and scatter as a function of ($\Gamma$, $M$, $z$) or ($M$, $z$)
\enddata
\tablecomments{Most \colossus modules (e.g., concentration and mass function) can automatically convert spherical overdensity mass definitions. There is, however, no simple conversion between friends-of-friends (FOF) and spherical overdensity (SO) definitions \citep{more_11_fof}.}
\end{deluxetable*}

The final quantity that relies on an integral of $E(z)$ is the linear growth factor, $D_+(z)$, the time-dependent normalization of the linear fluctuations in the density field. The value of $D_+$ is influenced by relativistic species at high redshift and by dark energy at low redshift, leading to different possibilities for the normalization of $D_+$ \citep[e.g.,][]{eisenstein_99, percival_05}. Internally, \colossus computes the growth factor such that $D_+(a) = a$ at high redshift, as is the case in an Einstein-de Sitter cosmology. By default, the growth factor is renormalized such that $D_+(z=0) = 1$, but the user can request either normalization.

We compute $D_+$ by splitting the redshift range into multiple segments. At $z > 10$, $D_+$ is approximated using Equation 5 in \citet{gnedin_11_dc},
\begin{align}
\label{eq:d1}
D_+(a) = & a + \frac{2}{3} a_{\rm mr} + \frac{a_{\rm mr}}{2 \ln(2) - 3} \nonumber \\
& \times \left[ 2 \sqrt{1+x} + \left( \frac{2}{3} + x \right) \ln{\frac{\sqrt{1+x} - 1}{\sqrt{1+x} + 1}} \right] \,,
\end{align}
where $x \equiv a/a_{\rm mr}$ and $a_{\rm mr} \equiv \Omega_{\rm rel,0} / \Omega_{\rm m,0}$ is the epoch of matter-radiation equality. If relativistic species are not included in the cosmology, $D_+(a) = a$ in this redshift regime. The relativistic corrections become very small at low redshift and can be neglected, but dark energy needs to be taken into account instead. The evolution of $D_+$ is determined by the differential equation
\begin{equation}
D_+'' + \frac{3}{2a} \left [ 1 - \frac{w(a)}{1 + X(a)} \right] D_+' - \frac{3 X(a)}{2 \left [1 + X(a) \right] a^2} D_+ = 0 \,,
\end{equation}
where $X(a) \equiv \Omega_{\rm m}(a) / \Omega_{\rm de}(a)$ \citep{linder_03_2}. Because $D_+(a)$ is approximately proportional to $a$, \citet{linder_03_2} suggest integrating $G_+ = D_+ / a$ so that
\begin{equation}
G_+'' + \left [ \frac{7}{2} - \frac{3 w(a)}{2 \left[ 1 + X(a) \right]} \right] \frac{G_+'}{a} + \frac{3 \left[ 1 - w(a)\right]}{2 \left [1 + X(a) \right] a^2} G_+ = 0 \,.
\end{equation}
We solve this equation by setting $G_+(a) = 1$ and $G_+'(a) = 0$ at $a = 10^{-3}$ and integrating forward using a Runge-Kutta integrator. For \LCDM cosmologies, we instead evaluate a simpler expression \citep{heath_77, peebles_80},
\begin{equation}
\label{eq:d2}
D_+(z) = \frac{5}{2} \Omega_{\rm m,0} E(z) \int_z^\infty \frac{1 + z'}{E(z')^3} dz' \,.
\end{equation}
The normalization of this expression corresponds to $D_+(a) = a$ in the matter-dominated regime to match the high-redshift formula of Equation~(\ref{eq:d1}). According to Equation~(\ref{eq:d2}), however, $E(z)$ does not include contributions from relativistic species, leading to a slight mismatch of $10^{-3}$ between Equations~(\ref{eq:d1}) and (\ref{eq:d2}) if relativistic species are included in the cosmology. To remove this disagreement, we interpolate $D_+$ in $\log(a)$ space between $z = 20$ and $z = 5$, leading to inaccuracies smaller than $3 \times 10^{-4}$ at any redshift. 

The growth factor calculation was tested against the iCosmos calculator ({\href{http://www.icosmos.co.uk}{icosmos.co.uk}) and found to agree to much better than a percent at low redshift (for \LCDM cosmologies). At high redshift, it is not clear that the two computations are directly comparable. 

\subsection{The Linear Matter Power Spectrum}
\label{sec:cosmo:pk}

Numerous important quantities in structure formation are based on the linear matter power spectrum, $P(k)$, the amplitude of density fluctuations as a function of scale. The power spectrum can be parameterized in terms of the primordial density field whose power spectrum is assumed to be a power law,
\begin{equation}
P(k) = T^2(k) \times k^{n_{\rm s}}
\end{equation}
where $T(k)$ is the transfer function and $n_{\rm s}$ is the scalar spectral index, a free parameter. A spectral index of one corresponds to a scale-free power spectrum in the sense that all modes contribute equal power when they enter the horizon, and that all modes contribute equally to fluctuations in the gravitational potential. Observationally, $n_{\rm s}$ is, indeed, measured to be close to unity (Table~\ref{table:cosmo}). The transfer function encapsulates the physics of growing and decaying perturbations, starting with the primordial power law and including effects such as the stagnation of growth during the radiation-dominated era, baryon acoustic oscillations (BAO), and various damping terms. After recombination, the evolution simplifies because both baryons and dark matter behave like pressureless fluids on large scales. 

The left top panel of Figure~\ref{fig:ps_sigma_xi} shows the power spectrum calculated using the Boltzmann code \textsc{Camb} \citep{lewis_00}. At small $k$, $T(k) = 1$, meaning that the power spectrum is equal to the primordial power law. The transfer function starts to decrease around the horizon scale at the epoch of matter-radiation equality. By definition, the linear power spectrum captures only the linear contribution to the growth of perturbations but not their nonlinear collapse. The time evolution of the linear component is described by the linear growth factor, $P(k,z) = D_+^2(z) P(k,0)$. The power spectrum is normalized to give a particular variance $\sigma_8 \equiv \sigma(8 \mpch, z = 0)$ (Section~\ref{sec:cosmo:sigma}).

While \colossus allows the user to supply a tabulated power spectrum, from, for example, numerical calculations using \textsc{Camb} or \textsc{Cmbfast} \citep{seljak_96}, the default is to compute $T(k)$ using the approximation of \citet[][see Table~\ref{table:models} for a listing of all fitting functions]{eisenstein_98}. This semi-analytical fitting function is accurate to better than 5\% if the effects of baryons are included (Figure~\ref{fig:ps_sigma_xi}). \colossus computes an interpolation table for the power spectrum when it is first evaluated for a given cosmology. This table covers wavenumbers between $10^{-20}$ and $10^{20} h {\rm Mpc}^{-1}$ and uses a non-uniform binning scheme with an increased density of bins near the BAO features. The interpolation accuracy is better than $2 \times 10^{-4}$ across the entire range of wavenumbers (bottom left panel of Figure~\ref{fig:ps_sigma_xi}).

\subsection{Variance}
\label{sec:cosmo:sigma}

Given the linear power spectrum, we can compute the variance of the density field
\begin{equation}
\label{eq:sigma}
\sigma^2(R,z) = \frac{1}{2 \pi^2} \int_0^{\infty} k^2 P(k,z) |\widetilde{W}(kR)|^2 dk \,,
\end{equation}
where $\tilde{W}$ is a filter. \colossus offers multiple options for this filter, namely the most commonly used top-hat in real space, 
\begin{equation}
\widetilde{W}_{\rm tophat} = \frac{3}{(kR)^3} \left[ \sin(kR) - kR \times \cos(kR) \right] \,.
\end{equation}
When this filter is used, $\sigma$ quantifies the variance in spheres of radius $R$. Alternative filters include a Gaussian,
\begin{equation}
\widetilde{W}_{\rm gaussian} = \exp \left[ \frac{-(kR)^2}{2} \right] \,,
\end{equation}
and a sharp k-space filter,
\begin{equation}
\widetilde{W}_{\rm sharp-k} = \Theta(1 - kR) \,,
\end{equation}
where $\Theta$ is the Heaviside step function. The variance grows with time according to the linear growth factor, $\sigma(R,z) = D_+(z)\sigma(R,0)$.

The center column of Figure~\ref{fig:ps_sigma_xi} compares $\sigma$ calculated from a numerically computed power spectrum and from the \citet{eisenstein_98} approximation. The agreement is better than $2\%$ at all relevant radii. The larger disagreement at very small $R$ is caused by two effects: first, the \citet{eisenstein_98} approximation overestimates the power at large $k$ because it ignores the pressure from baryons, and second, the \textsc{Camb} power spectrum can only be computed to a finite wavenumber, meaning that $\sigma$ is underestimated at small $R$.

The interpolation table for the variance covers a radial range from $10^{-12}$ to $10^3 \mpch$. The integration accuracy is set to $3 \times 10^{-3}$ or better, and the interpolation is accurate to $5 \times 10^{-3}$ or better (center bottom panel of Figure~\ref{fig:ps_sigma_xi}).

\vspace{1cm}

\subsection{Correlation Function}
\label{sec:cosmo:xi}

The linear matter--matter correlation function is given by yet another integral over the power spectrum, 
\begin{equation}
\label{eq:xi}
\xi_{\rm mm}(R, z) = \frac{1}{2 \pi} \int_0^{\infty} k^2 P(k,z) \frac{\sin(kR)}{kR} dk \,.
\end{equation}
This integral converges slowly because of the fast frequency and slow fall-off of the sinc term at high $kR$, making efficient interpolation particularly important. \colossus reduces the computation time by using two numerical tricks. First, the integrand is exponentially suppressed at $kR > 1000$ because those fast oscillations contribute a negligible amount to the overall integral. Second, Clenshaw--Curtis integration speeds up the integration of the sinusoidal term. 

The integration accuracy is set to an error of at most $10^{-5}$. The inaccuracy due to the power spectrum approximation is much larger, up to about 4\% between $10^{-2} \mpch$ and the zero-crossing of the correlation function (Figure~\ref{fig:ps_sigma_xi}). Around the zero-crossing, the relative error grows because the absolute value of the function becomes small. The interpolation for the correlation function spans radii between $10^{-3}$ and $500 \mpch$, and is accurate to better than 1\% (bottom right panel of Figure~\ref{fig:ps_sigma_xi}). 


\section{The Large-scale Structure Module}
\label{sec:lss}

\begin{figure*}
\centering
\includegraphics[trim = 5mm 3mm 2mm 1mm, clip, scale=0.68]{\figdir/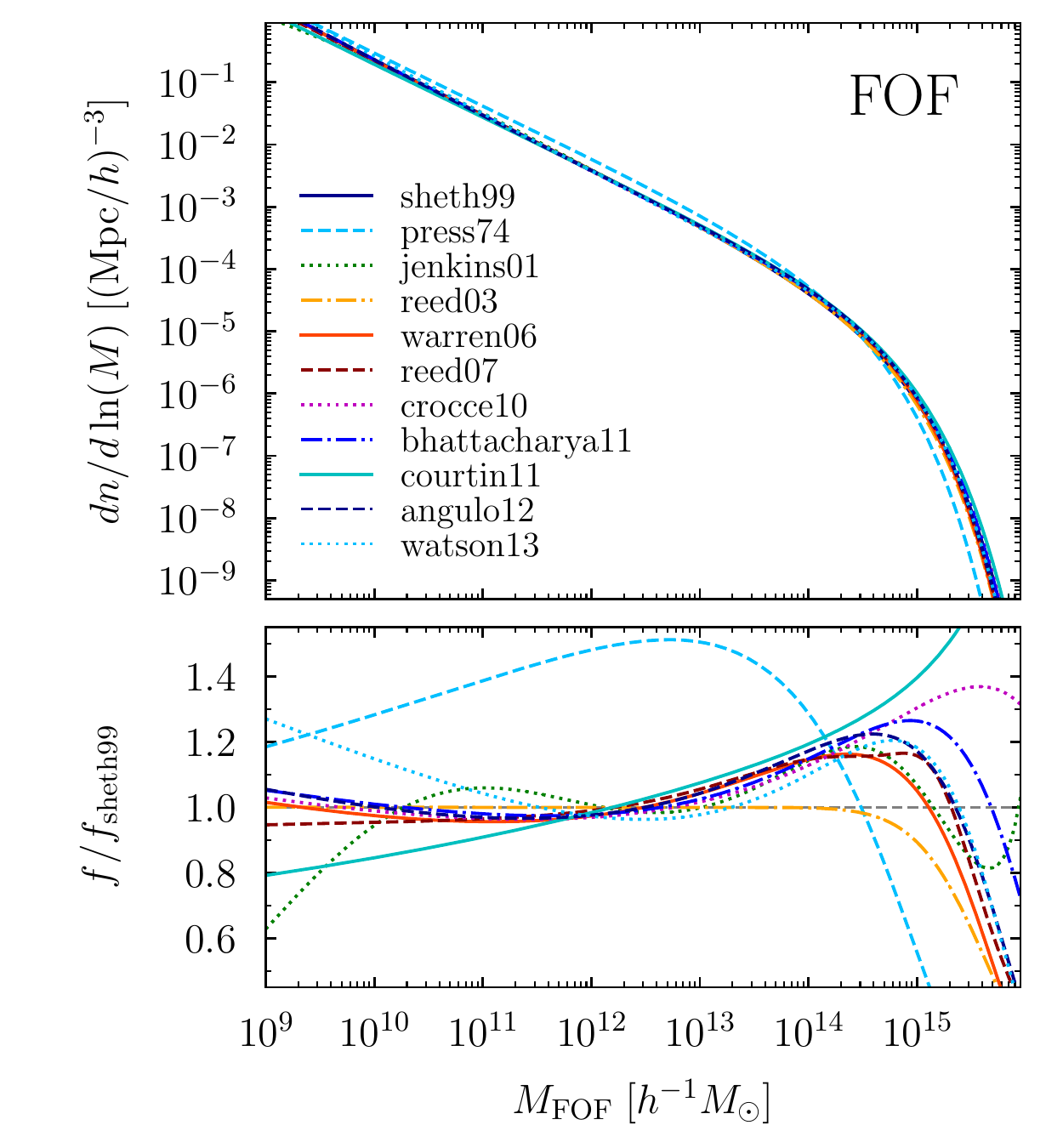}
\includegraphics[trim = 0mm 3mm 2mm 1mm, clip, scale=0.68]{\figdir/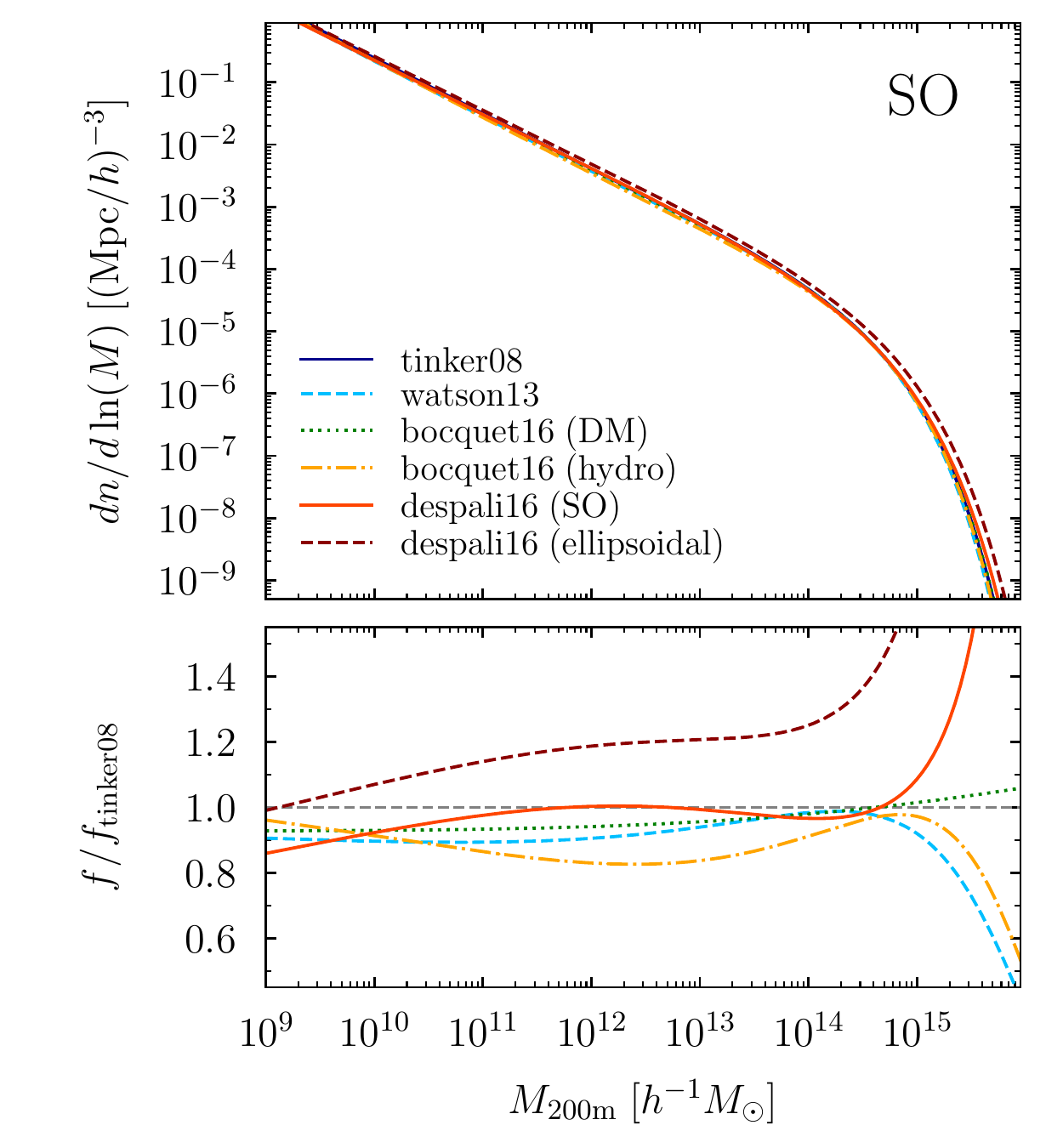}
\caption{Comparison of the halo mass function models implemented in \colossus, evaluated at $z = 0$ and for the planck15 cosmology. The left panel shows friends-of-friends mass functions, the right panel SO mass functions. The top panels demonstrate that the differences between the models are subtle, at least when viewed over a large range of halo mass. The bottom panels show the residual of the models with respect to the model of \citealt{sheth_99} (FOF) and \citealt{tinker_08} (SO). All FOF calibrations are based on a linking length of $0.2$ except for \citet{jenkins_01}, who used $0.164$. The SO mass functions are shown for the $\mtom$ mass definition, the differences between the models tend to increase toward higher overdensities. The \citet{despali_16} model was additionally fit to mass functions found by an ellipsoidal halo finder, meaning that some disagreement with the conventional SO mass functions is expected. The difference between the \citet{bocquet_16} calibrations for their dark matter-only and hydrodynamical simulations gives a hint as to the impact of baryons on the mass function.}
\label{fig:mfunc}
\end{figure*}

The large-scale structure (LSS) module covers the linear and nonlinear collapse of Gaussian random fields, including density peaks, their peak height and curvature, as well as the abundance of collapsed peaks (the halo mass function) and bias. Functions that deal with matter in general (as opposed to collapsed peaks) are based in the cosmology module, and functions that deal with the mass, radius, or structure of collapsed peaks are based in the halo module. All fitting functions implemented in the LSS module are listed in Table~\ref{table:models}.

\subsection{Peak Height, Peak Curvature, and Nonlinear Mass}
\label{sec:lss:peaks}

For the computations in this section, we assume that matter follows a linear Gaussian overdensity field $\delta$. In such a field, the statistical significance of a peak can be quantified by its ``peak height,'' $\nu = \delta / \sigma$, where $\sigma$ is the variance of the field. Halos, however, are nonlinearly collapsed objects, meaning that it is not obvious how to define their statistical significance compared to the linear density field. To construct an equivalent overdensity and variance, we translate some measure of halo mass (e.g, $\mvir$) into its Lagrangian radius, $R_{\rm L}$, the comoving radius of a sphere that encompasses the halo's mass at the mean density of the universe,
\begin{equation}
\label{eq:mtor}
M_{\rm L} = (4 \pi/3) \rho_{\rm m}(z=0) R_{\rm L}^3 \,.
\end{equation}
The peak height is then derived by comparing the variance on this radial scale with the overdensity above which density fluctuations are expected to collapse into halos,
\begin{equation}
\label{eq:nu}
\nu \equiv \frac{\delta_{\rm c}}{\sigma(M_{\rm L}, z)} = \frac{\delta_{\rm c}}{\sigma(M_{\rm L}, z = 0) \times D_+(z)} \,.
\end{equation}
The critical overdensity for collapse,
\begin{equation}
\delta_{\rm c,EdS} = \frac{3}{5} \left( \frac{3 \pi}{2} \right)^{2/3} \simeq 1.68647 \,,
\end{equation}
is derived from the spherical top-hat collapse model in an Einstein-de Sitter universe \citep{gunn_72}. In other cosmologies, small corrections apply, namely 
\begin{equation}
\label{eq:deltac_z1}
\delta_{\rm c}(z) \simeq \delta_{\rm c,EdS} \Omega_{\rm m}(z)^{0.0185}
\end{equation}
in non-flat cosmologies without dark energy and 
\begin{equation}
\label{eq:deltac_z2}
\delta_{\rm c}(z) \simeq \delta_{\rm c,EdS} \Omega_{\rm m}(z)^{0.0055}
\end{equation}
in flat cosmologies with dark energy \citep{mo_10_book}. These corrections change $\delta_{\rm c}$ by less than one percent for realistic cosmologies, and \colossus applies them only if requested by the user. Finally, the nonlinear mass, $M^*$, is defined as the mass where $\sigma(M^*) = \delta_{\rm c}$, and thus $\nu(M^*) = 1$.

By analogy with the peak height, we can define the curvature of a field as $x \equiv -\nabla^2 \delta/\sigma_2$, where $\sigma_2$ is the second moment of the variance (Equation~4.6(c) in \citealt{bardeen_86} or Equation~\ref{eq:sigma} with a factor of $k^4$ inside the integral). For halos, peak curvature is a measure of their steepness, though other definitions exist \citep{dalal_08, dalal_10}. As with peak height, we need to define a measure that applies to nonlinearly collapsed objects. \citet{bardeen_86} derived an average curvature as a function of peak height, $\avg{x}$, which can be computed by integrating their Equations~(A14) and (A15). They also give a 1\% accurate fitting function in Equation~(6.13); both versions are available in \colossus. The higher-order moments of the variance (such as $\sigma_2$) must be computed using a Gaussian filter rather than a top-hat because the integral does not converge in the latter case. 

One important issue with peak curvature is the cloud-in-cloud problem: while $\avg{x}$ gives the average curvature of peaks of a certain significance, not all of those peaks end up forming halos because some of them are absorbed into other, larger peaks. Thus, $\avg{x}$ does not necessarily correspond to the average curvature of the peaks that create halos of a particular mass \citep{bardeen_86}.

\subsection{Halo Mass Function}
\label{sec:lss:mfunc}

The halo mass function quantifies how many halos of a given mass have formed at a given redshift and cosmology. According to the Press--Schechter ansatz \citep{press_74, bond_91}, the mass function is expected to be universal (i.e., independent of redshift and cosmology) when expressed as the multiplicity function, $f(\sigma)$. This function translates to the number of halos per logarithmic mass interval as
\begin{equation}
\frac{dn}{d \ln(M)} = f(\sigma) \frac{\rho_{\rm m,0}}{M} \frac{d \ln(\sigma^{-1})}{d \ln(M)} \,,
\end{equation}
where $\sigma(M)$ is the variance on the Lagrangian scale of a halo as defined in Equation~\ref{eq:sigma}. The multiplicity function can be interpreted as the fraction of mass that has collapsed to form halos in a unit interval of $\ln(\sigma^{-1})$. \colossus can return the mass function in units of $f(\sigma)$, in the number density per logarithmic interval in mass, $dn/d\ln(M)$, or as the dimensionless quantity $M^2 / \rho_{\rm m,0} dn/dM$. \citet{press_74} derived the generic prediction that
\begin{equation}
f_{\rm PS}(\sigma) = \sqrt{\frac{2}{\pi}} \frac{\deltac}{\sigma} \exp \left(-\frac{\deltac^2}{2 \sigma^2} \right) \,,
\end{equation}
but this form was found to be in disagreement with numerical simulations, leading to numerous improved fitting functions for $f(\sigma)$. The models implemented in \colossus are listed in Table~\ref{table:models}, Figure~\ref{fig:mfunc} shows a comparison for halos defined via the friends-of-friends algorithm \citep[FOF,][]{davis_85}, and for halos defined via the spherical overdensity (SO) definition.

While the universality of $f(\sigma)$ is still debated \citep[e.g.,][]{tinker_08, bhattacharya_11}, its redshift evolution is agreed to be relatively mild. Some models encode an evolution explicitly, while some exhibit a slightly changing $f(\sigma)$ due to the weak redshift evolution of $\deltac$ (Equations~(\ref{eq:deltac_z1}) and (\ref{eq:deltac_z2})). The user can choose whether this evolution is taken into account or not (unless the model explicitly specifies that $\deltac$ should be a particular constant, e.g. \citealt{courtin_11}). Finally, models for the SO mass function that can rescale between different mass definitions introduce a redshift dependence simply because of the dependence of the overdensity threshold on redshift \citep[e.g.,][]{tinker_08, watson_13_mf, despali_16}. 

The mass functions from \colossus were compared to the \textsc{hmf} package \citep{murray_13} and exhibit excellent agreement. However, some choices (such as the definition of the collapse overdensity) differ in the default versions of the two codes.

\subsection{Halo Bias}
\label{sec:lss:bias}

Halo bias quantifies the excess clustering of collapsed halos over that of dark matter. Thus, bias can be defined as the ratio of the halo and linear matter power spectra, 
\begin{equation}
b \equiv \sqrt{\frac{P_{\rm halo}(k)}{P_{\rm lin,matter}(k)}} \,.
\end{equation}
The bias is, in principle, a function of both halo mass and scale, but is expected to become scale-independent at large radii \citep[e.g.,][]{sheth_99, tinker_05, smith_07}. Thus, all bias models implemented in \colossus ignore the scale dependence and quantify the large-scale bias. A simple prediction for the bias can be derived from the peak-background split ansatz \citep{cole_89, mo_96}. This model was modified at low masses by \citet{jing_98} who calibrated their model on simulations of scale-free cosmologies. \citet{sheth_01} further improved upon the \citet{mo_96} prescription by taking the ellipsoidal nature of the collapse into account. \citet{tinker_10} undertook a careful numerical calibration and included a prescription for the dependence of bias on the halo mass definition. As a result, their model exhibits a slight redshift dependence for most mass definitions. In addition, a number of numerical calibrations of bias have been undertaken \citep[e.g.,][]{manera_10, pillepich_10}. Figure~\ref{fig:bias} compares the bias models implemented in \colossus as a function of peak height.

\begin{figure}
\centering
\includegraphics[trim = 2mm 3mm 2mm 3mm, clip, scale=0.69]{\figdir/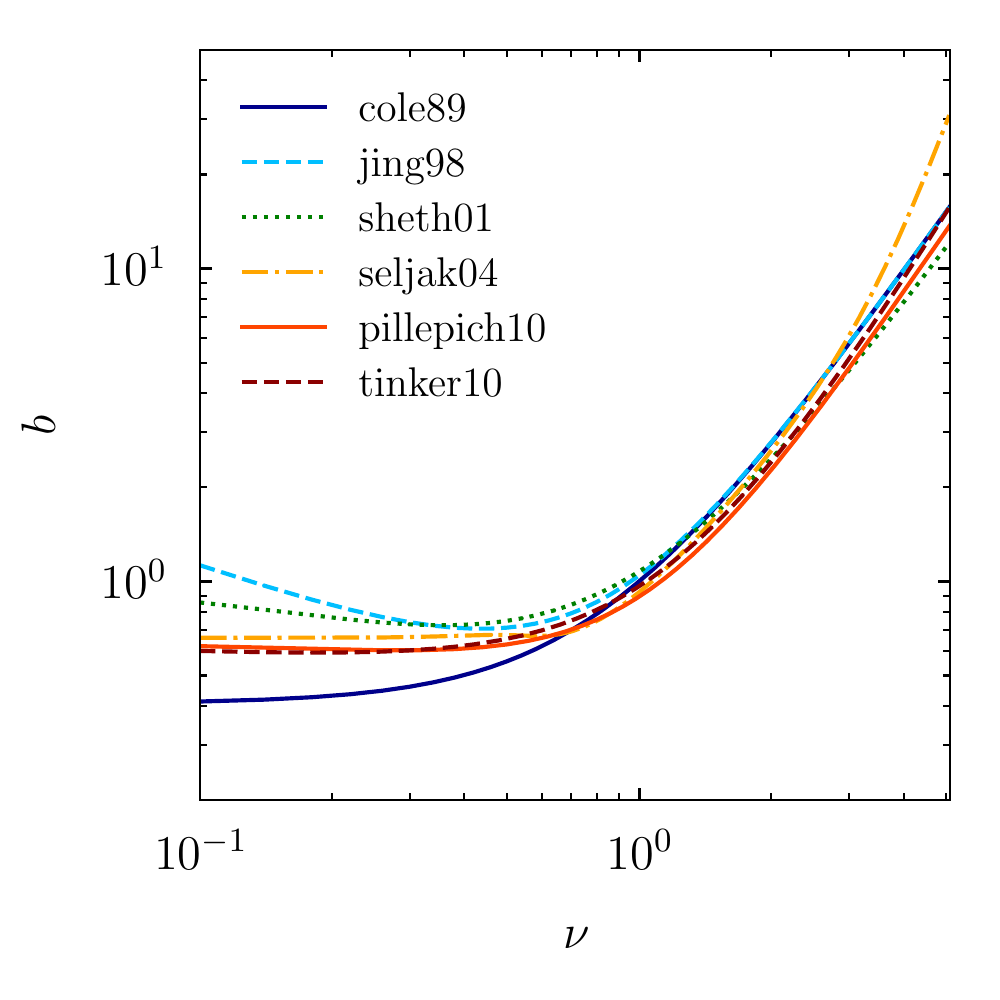}
\caption{Comparison of the halo bias models implemented in \colossus, computed for the planck15 cosmology. The definition of peak height used in the model calibrations can vary slightly, but such differences should have a negligible impact on the model predictions. The \citet{seljak_04} model is shown without the cosmological correction term in their Equation (6), and the \citet{tinker_10} model is shown for the $\mtom$ mass definition.}
\label{fig:bias}
\end{figure}


\section{The Halo Module}
\label{sec:halo}

\begin{figure*}
\centering
\includegraphics[trim = 0mm 0mm 2mm 0mm, clip, scale=0.68]{\figdir/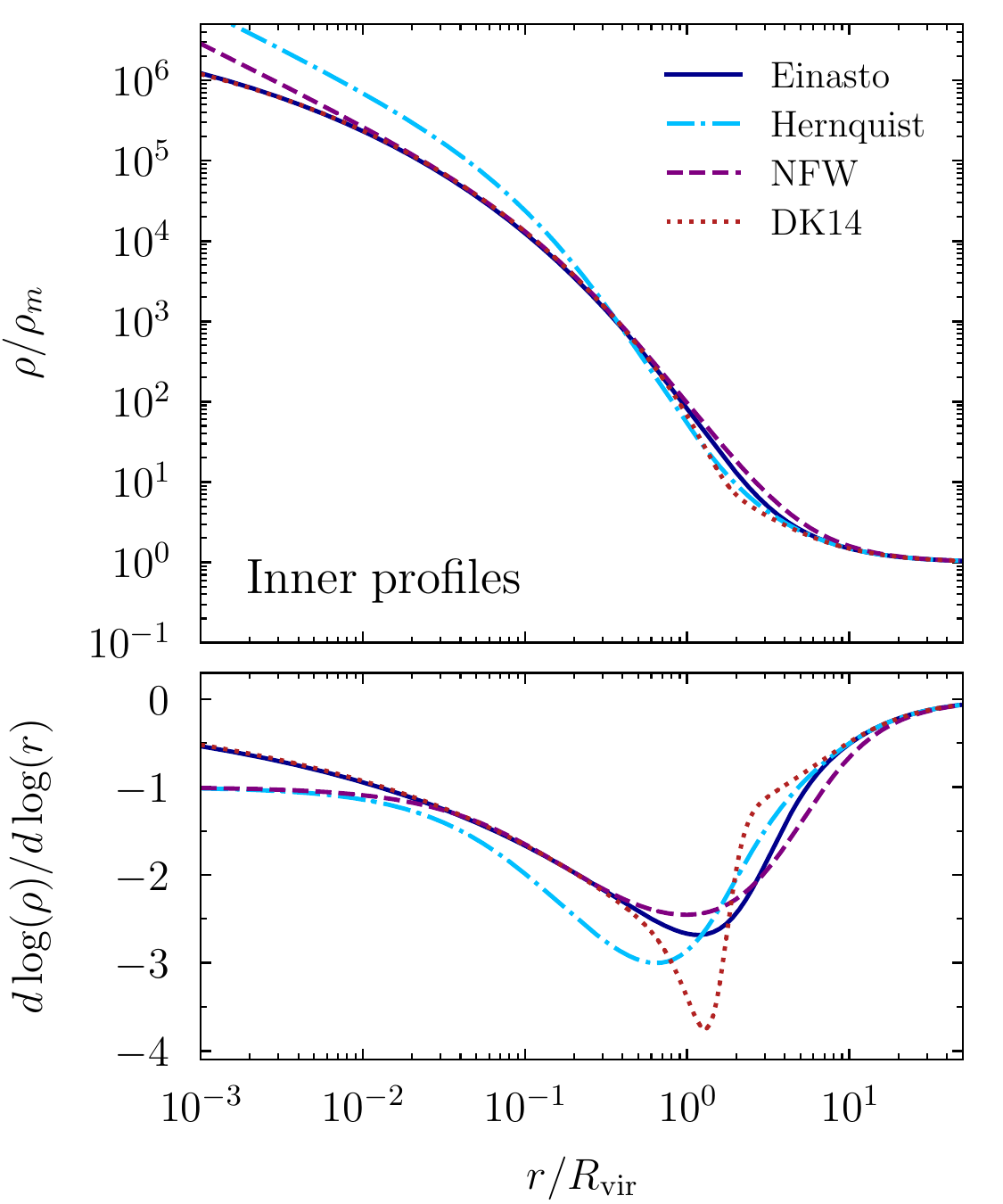}
\includegraphics[trim = 20mm 0mm 2mm 0mm, clip, scale=0.68]{\figdir/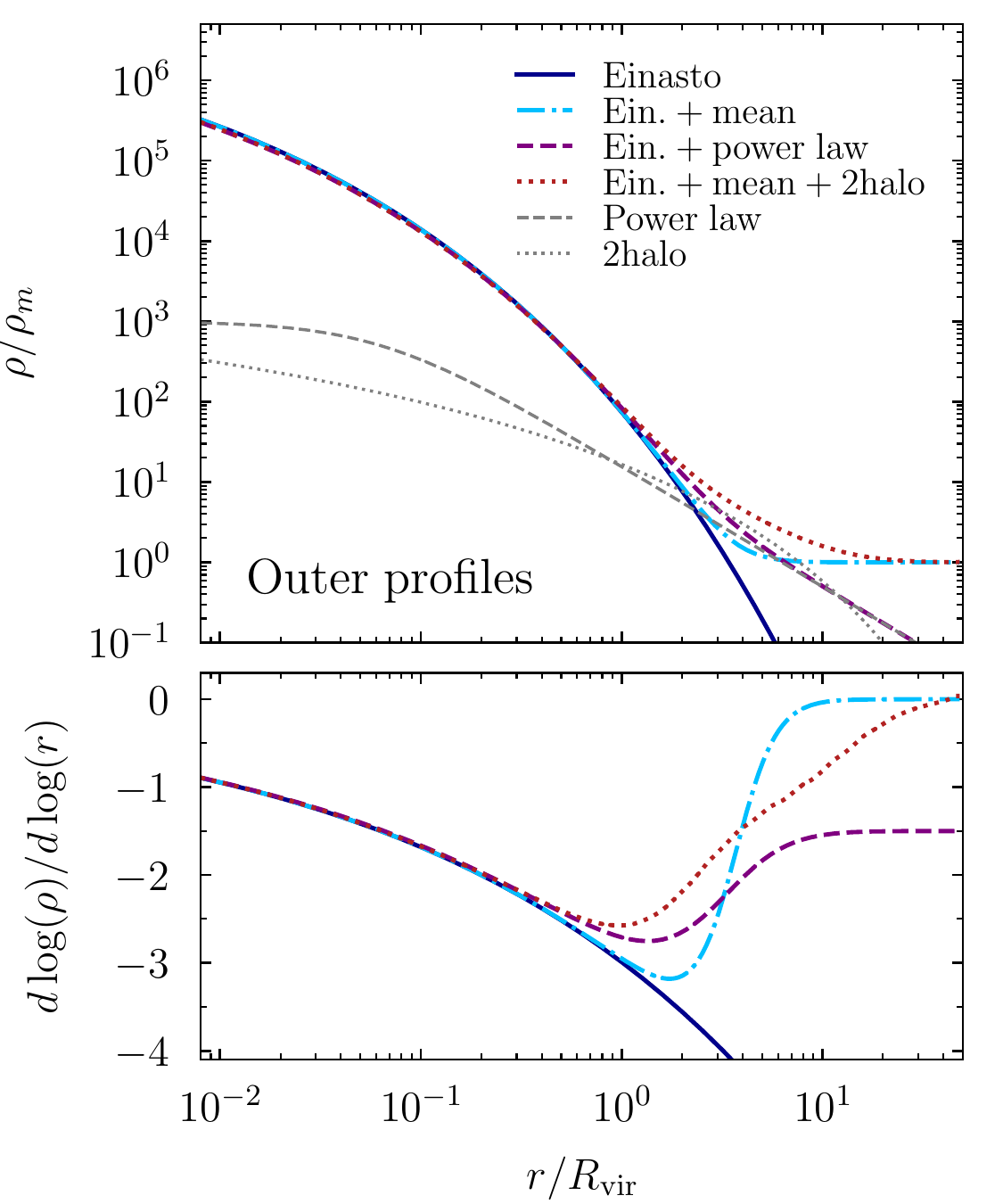}
\caption{Halo density profiles (top panels) and their logarithmic slopes (bottom panels). All profiles correspond to a halo with $\mvir = 10^{15} \msunh$ and $c_{\rm vir} = 5$ at $z = 0$. Left: a comparison of the Einasto, Hernquist, NFW, and DK14 profile forms. For realism and comparability, a power-law outer profiles, as well as the mean density of the universe, were added to all profiles at large radii. Right: a comparison of the outer profile terms available in \colossus. The gray lines show the density contributions due to the power law and correlation-function outer terms. The power-law term is cut off at a particular maximum density in order to avoid spurious contributions at very small radii.}
\label{fig:profiles}
\end{figure*}

The halo module is concerned with the spherically averaged structure of dark matter halos and their boundaries. Table~\ref{table:models} gives an overview of the fitting functions implemented for halo density profiles, concentration, and the splashback radius.

\subsection{Spherical Overdensity}
\label{sec:halo:so}

The most commonly used definition of a halo's boundary and mass is the SO definition where the radius is defined to enclose a particular overdensity $\Delta$ such that
\begin{equation}
\label{eq:so}
\mdelta = \frac{4\pi}{3} \Delta \rho_{\rm ref} \rdelta^3 \,,
\end{equation}
where $\rho_{\rm ref}$ is either the critical or mean matter density of the universe \citep[e.g.][]{cole_96},
\begin{equation}
\label{eq:mdelta_m}
M_{\Delta \rm m} = M(<R_{\Delta \rm m})= \frac{4 \pi}{3} \Delta\rho_{\rm m}(z)R^3_{\Delta {\rm m}} \,,
\end{equation}
for example, $\rtom$ and $\mtom$, or 
\begin{equation}
\label{eq:mdelta_c}
M_{\Delta \rm c} = M(<R_{\Delta \rm c})= \frac{4 \pi}{3} \Delta\rho_{\rm c}(z)R^3_{\Delta {\rm c}} \,,
\end{equation}
for example, $\rtoc$ and $\mtoc$. The labels $\mvir$ and $\rvir$ indicate a varying overdensity $\Delta_{\rm vir}(z)$, which \colossus computes using the approximation of \citet{bryan_98}. \colossus offers a number of basic routines related to SO masses and radii, beginning with the computation of the density threshold $\rho_{\rm ref}$. Based on this threshold, we define a typical velocity 
\begin{equation}
v_{\Delta} \equiv \sqrt{\frac{GM_{\Delta}}{R_{\Delta}}}
\end{equation}
which we use to define the dynamical time of the halo as the time it takes to cross $2R_{\Delta}$,
\begin{equation}
\label{eq:tdyn}
t_{\rm dyn}(z) \equiv t_{\rm cross}(z) = \frac{2 \rdelta}{\vdelta} = 2^{3/2} t_{\rm H}(z) \left( \frac{\rhodelta(z)}{\rhoc(z)} \right)^{-1/2} \,.
\end{equation}
Notably, this time does not depend on the distribution of matter inside the halo or even the halo radius, but only on $\rhodelta$ and the Hubble time,
\begin{equation}
\label{eq:thubble}
t_{\rm H}(z) \equiv \frac{1}{H(z)} = \sqrt{\frac{3}{8 \pi G \rhoc(z)}} \,.
\end{equation}
Alternatively, \colossus offers the time to pericenter (traveling one $\rdelta$) or the orbital time (traveling $2 \pi \rdelta$) as definitions of the dynamical time.

If we wish to convert between $R_{\Delta}$ and $M_{\Delta}$ for different overdensity definitions, the results depend on the density profile. In particular, we need to solve Equation~(\ref{eq:so}) numerically given a particular $\rho_{\rm ref}$ and $M(r)$. By default, \colossus uses a Navarro--Frenk--White (NFW) profile (see Section~\ref{sec:halo:profiles}) to convert between definitions, but the user can also choose other profile models. The concentration can either be given by the user or be automatically computed using a model of the concentration--mass relation (Section~\ref{sec:halo:concentration}). 

The conversion between different overdensity definitions constitutes a special case of a more general conversion where not only the overdensity is varied but also the redshift. Here, the assumption is that the halo density profile is static in time, but that the spherical overdensity ``pseudo-evolves'' due to the change in the critical or mean density \citep[see, e.g.,][]{diemand_05, cuesta_08, diemer_13_pe, zemp_14, more_15}. This more general routine is also implemented in \colossus. All conversion routines are based on the \citet{brent_73} root finding algorithm with a default accuracy of $10^{-12}$.

\subsection{Density Profiles}
\label{sec:halo:profiles}

The density profiles of dark matter halos have been studied extensively in the literature, generally based on the results of numerical simulations. All profile models implemented in \colossus describe the spherically averaged density profile $\rho(r)$. The implementation of each model relies on a base class that numerically computes numerous quantities, including:
\begin{itemize}
\item the linear and logarithmic derivatives of density, $d \rho / dr$ and $d \ln(\rho) / d \ln(r)$;
\item the enclosed mass, $M(<r)$;
\item the surface density $\Sigma$ and excess surface density $\Delta \Sigma$ (sometimes referred to as ``differential surface density''), defined as $\Delta \Sigma(R) \equiv \Sigma(<R) - \Sigma(R)$ where $\Sigma(<R)$ is the average surface density inside $R$ weighted by area,
\begin{equation}
\Sigma(<R) \equiv \frac{2}{R^2} \int_0^{R} r \Sigma(r) dr \,;
\end{equation}
\item the circular velocity, $v_{\rm c}(r) = \sqrt{GM(<r)/r}$, and the maximum circular velocity, $v_{\rm max}$; and
\item SO radius and mass for a given overdensity definition and redshift.
\end{itemize}
A profile model is fully specified by only the density $\rho(r)$. The quantities listed above are, by default, computed numerically from $\rho(r)$ unless a model implementation overwrites them when, for example, analytical solutions are available. In \colossus, the density profile is modeled as the sum of an ``inner'' profile (or 1-halo term) as well as an arbitrary number of ``outer'' profiles. The left panels of Figure~\ref{fig:profiles} show a comparison of the inner profile models implemented in \colossus.
\begin{itemize}
\item The three-parameter profile of \citet{einasto_65} is described by a logarithmic slope that changes progressively with radius, 
\begin{equation}
\rho_{\rm Einasto}(r) = \rho_{\rm s} \exp \left( -\frac{2}{\alpha} \left[ \left( \frac{r}{r_{\rm s}} \right)^\alpha -1 \right] \right) \,.
\end{equation}
The user can either choose the shape parameter $\alpha$ or let it be determined by the fitting function of \citet{gao_08},
\begin{equation}
\label{eq:gao}
\alpha(\nu) = 0.155 + 0.0095 \nu^2 \,,
\end{equation}
which gives $\alpha = 0.25$ for the massive halo shown in Figure~\ref{fig:profiles}. 

\item The two-parameter \citet{hernquist_90} profile is given by the expression
\begin{equation}
\rho_{\rm Hernquist}(r) = \frac{\rho_{\rm s}}{\left( \frac{r}{r_{\rm s}} \right)\left( 1 + \frac{r}{r_{\rm s}} \right)^3} \,.
\end{equation}
This profile approaches power laws of slope $-1$ and $-4$ at small and large radii, respectively, turning over around the scale radius $r_{\rm s}$.

\item The NFW profile \citep{navarro_95, navarro_96, navarro_97} changes the outer slope of the Hernquist profile to $-3$,
\begin{equation}
\rho_{\rm NFW}(r) = \frac{\rho_{\rm s}}{\left( \frac{r}{r_{\rm s}} \right)\left( 1 + \frac{r}{r_{\rm s}} \right)^2} \,.
\end{equation}

\item In order to account for the steepening at the splashback radius (Section~\ref{sec:halo:rsp}), \citet[][DK14]{diemer_14} combined the Einasto profile at small radii with a steepening function in the outer profile,
\begin{equation}
\rho_{\rm DK14}(r) = \rho_{\rm Einasto} \times \left[ 1 + \left( \frac{r}{r_{\rm t}}\right)^\beta \right]^{-\frac{\gamma}{\beta}} \,,
\end{equation}
where $\beta$ determines how rapidly this steepening happens and $\gamma$ represents the limiting slope of the steepening term. \citet{diemer_14} recommend $(\beta, \gamma) = (4, 8)$ or $(6, 4)$, depending on how the halo sample was selected (the profile shown in Figure~\ref{fig:profiles} uses $(4, 8)$). The turnover radius $r_{\rm t}$ depends on the location of the splashback radius and thus on the mass accretion rate. The DK14 profile makes sense only when combined with a prescription for the outer profile as described below.
\end{itemize}
Regardless of the parameterization of the given profile, the user can initialize a profile from a mass and concentration that are automatically converted to the native parameters (e.g., $\rho_{\rm s}$ and $r_{\rm s}$). Furthermore, \colossus provides an arbitrary spline-interpolated profile based on a table of either $\rho(r)$ or $M(r)$. Care needs to be taken when integrating over such profiles if, for example, the radial extent of the table is not sufficient to compute the surface density.

The left panels of Figure~\ref{fig:profiles} show a comparison of these profile forms and their logarithmic derivatives. All models of the inner profile become somewhat unrealistic at $r \gsim \rvir$ where the outer profile begins to contribute significantly. Physically, the excess density at large radii is due to a combination of the nonlinear infall of matter into the halo and the statistical contribution from neighboring halos (the 2-halo term, e.g. \citealt{smith_03_powerspec}; \citealt{hayashi_08}). In \colossus, these contributions are modeled as the sum of an arbitrary combination of the following terms: 
\begin{itemize}
\item The mean density of the universe, $\rho = \rho_{\rm m}(z)$. This term should always be included if the profile is evaluated at large radii.
\item An estimate of the 2-halo term based on the linear matter--matter correlation function,
\begin{equation}
\rho_{\rm 2h}(r, z) = \rho_{\rm m}(z) \xi(r, z) b(\nu) \,,
\end{equation}
where the bias $b(\nu)$ can be estimated based on a model of halo bias (Section~\ref{sec:lss:bias}). The 2-halo term shown in the right panels of Figure~\ref{fig:profiles} corresponds to $b = 6.1$, appropriate for a very massive halo.
\item A power-law outer profile \citep[e.g.,][]{diemer_14} that can be used to approximate the profile of infalling matter \citep{bertschinger_85} or mimic a 2-halo term. Mathematically the power-law outer term is described as
\begin{equation}
\rho_{\rm PL}(r, z) = \rho_{\rm m}(z) \frac{a}{\frac{1}{\rho_{\rm max}} + \left( \frac{r}{r_{\rm pivot}} \right)^b} \,,
\end{equation}
where $a$ is a normalization, $b$ is the slope, and $r_{\rm pivot}$ is an arbitrary pivot radius that can be set to either a fixed radius or an SO radius. The limiting density $\rho_{\rm max}$ is introduced to avoid spurious contributions at small radii. The power-law outer profile shown in Figure~\ref{fig:profiles} has $a = 1$ and $b = 1.5$ \citep{diemer_14}.
\end{itemize}
Finally, the density profile object provides powerful functionality to fit profile models to data. The data can be density, enclosed mass, or (excess) surface density. Estimates of the uncertainties on those data and their covariances are taken into account if desired, and the fit can be performed using either a least-squares or MCMC algorithm \citep{goodman_10}.

\subsection{Concentration}
\label{sec:halo:concentration}

\begin{figure}
\centering
\includegraphics[trim = 8mm 8mm 2mm 3mm, clip, scale=0.69]{\figdir/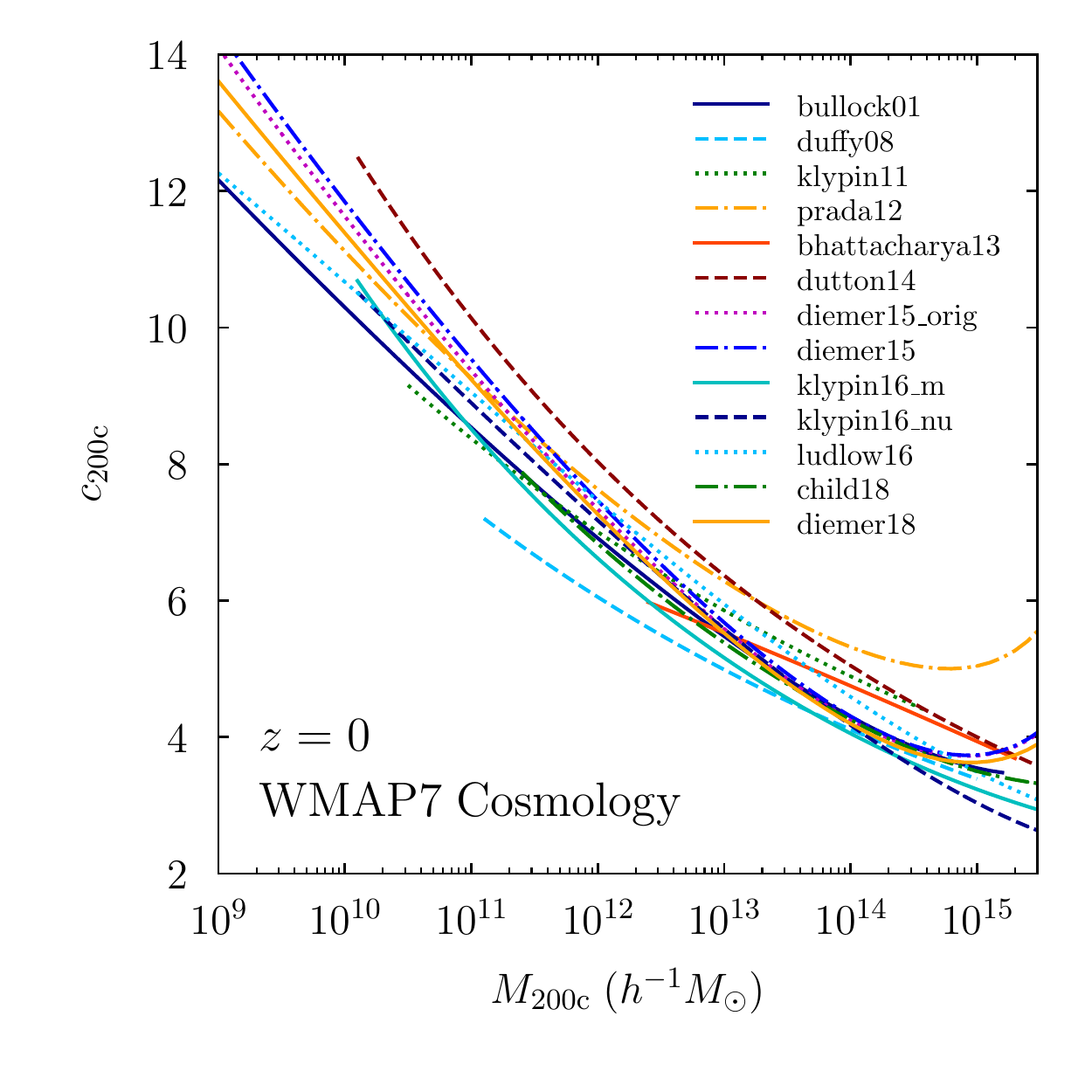}
\caption{Halo concentration at $z = 0$ for the bolshoi (WMAP7) cosmology (Table~\ref{table:cosmo}), as predicted by the various concentration models implemented in \colossus.}
\label{fig:concentration}
\end{figure}

\begin{figure}
\centering
\includegraphics[trim = 6mm 8mm 2mm 3mm, clip, scale=0.69]{\figdir/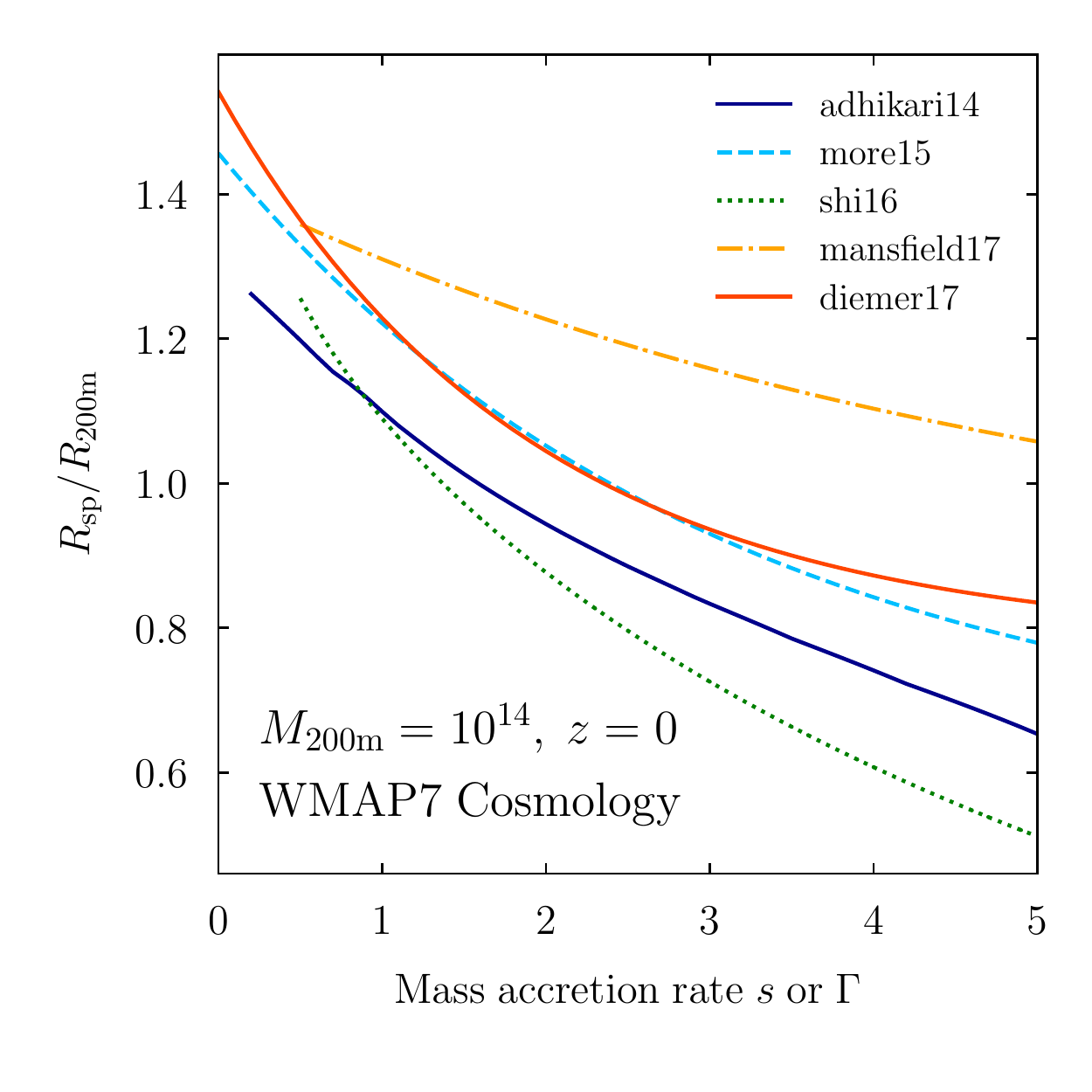}
\caption{Model predictions for the splashback radius as a function of mass accretion rate, evaluated for a halo of mass $\mtom = 10^{14} \msunh$ at $z = 0$ in the bolshoi (WMAP7) cosmology (Table~\ref{table:cosmo}). This comparison is complicated by the different definitions of the accretion rate adopted by the models.}
\label{fig:splashback}
\end{figure}

As illustrated in the previous section, the most popular expressions for halo density profiles are characterized by a scale radius, $r_{\rm s}$. Such profiles are more naturally described by an SO mass and a concentration, the ratio between the SO radius and the scale radius, $c_{\Delta} \equiv R_{\Delta} / r_{\rm s}$. If concentration can be quantified as a function of mass and other parameters, a full description of the halo profile can be derived from only an input mass, making the concentration--mass relation a paramount tool in halo modeling. However, the c--M relation turns out to depend on redshift and cosmology in a non-trivial fashion, and to exhibit significant halo-to-halo scatter. Thus, numerous models for the c--M relation have been proposed in the literature. Most works rely on NFW-based concentrations, i.e. fitting halo profiles with the NFW form to derive $r_{\rm s}$, but other parameterizations exist (e.g., \citealt{klypin_11}, \citealt{prada_12}; see \citealt{dutton_14} for a comparison with Einasto-based concentrations).

The \colossus concentration module implements several models for the c--M relation (Table~\ref{table:models}), Figure~\ref{fig:concentration} shows a comparison of their predictions at $z = 0$. The models broadly fall into three categories. First, the c--M relation is reasonably well described by a power law for a given redshift and cosmology, and within a certain range of halo mass \citep{duffy_08, klypin_11, dutton_14, klypin_16, child_18}. Second, concentration is strongly correlated with halo age, giving rise to models that base concentration on a description of halo formation times \citep[e.g.,][]{bullock_01, wechsler_02, zhao_03_concentration, zhao_09, ludlow_16}. Finally, the c--M relation is almost universal with redshift when mass is expressed as peak height (Section~\ref{sec:lss:peaks}), leading to models that parameterize the dependence either empirically or based on physical arguments \citep{prada_12, bhattacharya_13, diemer_15, diemer_18_cm}. Some noteworthy models are not implemented in \colossus, most commonly because they are not based on simple analytical expressions and thus demand significant computation \citep[e.g.][]{eke_01, zhao_09, correa_15_c}.

A further complication arises because the models quantify different definitions of concentration, including $\ctom$, $\ctoc$, and $\cvir$. The conversion to the desired mass definition is performed automatically by \colossus and can be based on either NFW or DK14 profiles. This process is iterative because an input mass in one definition may lead to an output concentration in another definition, which in turn changes the conversion between the masses. The conversion can introduce slight inaccuracies into the predictions for concentration because the profile models do not describe real halos perfectly \citep{diemer_15}. Moreover, the c--M relations measured in simulations depend on technical details such as the fitting procedure and binning \citep{dooley_14, meneghetti_14}. Taking these effects into account would add significant complication and is beyond the scope of \colossus.

\subsection{The Splashback Radius}
\label{sec:halo:rsp}

The splashback radius, $\rsp$, has recently been proposed as a physically motivated definition of the halo boundary \citep{diemer_14, adhikari_14, more_15} and has since been observed in stacked cluster density profiles \citep{more_16, baxter_17, chang_18}. Unlike SO radii, the splashback depends on the dynamical state of a halo, namely on its mass accretion rate. A number of models for this dependence, and additional dependencies on halo mass and redshift, have been proposed (Table~\ref{table:models}). \colossus provides a general function to evaluate the model predictions for the splashback radius, splashback mass, enclosed overdensity, and the scatter in those quantities. Figure~\ref{fig:splashback} compares the model predictions for the splashback radius of halo with $\mtom = 10^{14} \msunh$ at $z = 0$.

The semi-analytical models of \citet{adhikari_14} and \citet{shi_16} are based on spherically collapsing shells whose radial trajectories are integrated numerically. While the \citet{adhikari_14} model predicts that $\rsp/\rtom$ should depend only on the mass accretion rate $s \equiv d \ln(M) / d \ln(a)$, the \citet{shi_16} model also predicts an evolution with redshift. Numerical calibrations have shown that $\rsp/\rtom$ does depend on redshift, regardless of the way $\rsp$ is measured: from stacked density profiles \citep{more_15}, non-spherical splashback shells \citep{mansfield_17}, or particle orbits \citep{diemer_17_sparta, diemer_17_rsp}. In these calibrations, the mass accretion rate is defined as 
\begin{equation}
\Gamma \equiv \frac{\Delta \ln(M)}{\Delta \ln(a)}
\end{equation}
because the instantaneous rate is not a well-defined quantity in simulations. The time interval over which the mass accretion rate is measured has also varied slightly between the different models but is generally close to a dynamical time. The different definitions of the mass accretion rate complicate the interpretation of the comparison in Figure~\ref{fig:splashback}.


\section{Future Development}
\label{sec:conclusions}

I have presented \colossus, an open-source python toolkit that is available at \href{https://bitbucket.org/bdiemer/colossus}{bitbucket.org/bdiemer/colossus}. I anticipate that the code will expand significantly over the coming years, and that this development will be driven by the needs of its users. Additions could take the form of new functionality (e.g., new fitting functions), new physics (e.g., warm dark matter), or entirely new modules (e.g., calculations related to galaxy formation). 

Thus, all \colossus users are encouraged to suggest changes and to use the issue tracking system on BitBucket to report bugs, unclear documentation, and feature requests. Most importantly, however, I invite collaborators! While \colossus has hitherto been essentially a single-developer project, I would like this situation to change in the future.


\vspace{0.5cm}

\colossus was born while I was working on my PhD thesis, and I am grateful to Andrey Kravtsov for his mentoring and support during that time, as well as for contributing his MCMC routine. I am also grateful to Matt Becker, whose \textsc{CosmoCalc} code was a great inspiration during the early development of \colossus. I thank all those who have tested \colossus and made suggestions, namely Douglas Applegate, Neal Dalal, Daniel Eisenstein, Lehman Garrison, Andrew Hearin, Wayne Hu, Michael Joyce, Andrey Kravtsov, Alexie Leauthaud and her students, Philip Mansfield, Tom McClintock, Surhud More, and Steven Murray. I am indebted to the referee, Frank van den Bosch, whose extremely careful reading of this paper caught numerous errors and inaccuracies. Furthermore, I thank Savvas Koushiappas for suggesting this paper and Sownak Bose for comments on a draft. \colossus makes extensive use of the numpy (\href{http://www.numpy.org/}{numpy.org}) and scipy (\href{https://scipy.org/}{scipy.org}) libraries. I gratefully acknowledge the financial support of an Institute for Theory and Computation Fellowship. Support for Program number  HST-HF2-51406.001-A was provided by NASA through a grant from the Space Telescope Science Institute, which is operated by the Association of Universities for Research in Astronomy, Incorporated, under NASA contract NAS5-26555. This research was supported in part by the National Science Foundation under Grant No. NSF PHY17-48958.


\bibliographystyle{aasjournal}
\bibliography{../../../_LatexInclude/sf.bib}

\end{document}

%% file: commands.tex
\newcommand{\mpch}{\>h^{-1}{\rm {Mpc}}}

\newcommand{\msunh}{\>h^{-1} M_\odot}

\def\gcm3{\mathrm{g} / \mathrm{cm}^3}

\def\LCDM{$\Lambda$CDM\xspace}

\def\mvir{M_{\rm vir}}
\def\rvir{R_{\rm vir}}
\def\cvir{c_{\rm vir}}

\def\mtom{M_{\rm 200m}}
\def\rtom{R_{\rm 200m}}
\def\ctom{c_{\rm 200m}}

\def\mtoc{M_{\rm 200c}}
\def\rtoc{R_{\rm 200c}}
\def\ctoc{c_{\rm 200c}}

\def\mdelta{M_{\Delta}}
\def\rdelta{R_{\Delta}}

\def\vdelta{v_{\Delta}}

\def\rhoc{\rho_{\rm c}}

\def\rhodelta{\rho_{\Delta}}
\def\deltac{\delta_{\rm c}}

\def\colossus{\textsc{Colossus}\xspace}

\def\gtsima{$\; \buildrel > \over \sim \;$}
\def\ltsima{$\; \buildrel < \over \sim \;$}
\def\prosima{$\; \buildrel \propto \over \sim \;$}
\def\gsim{\lower.7ex\hbox{\gtsima}}
\def\lsim{\lower.7ex\hbox{\ltsima}}
\def\simgt{\lower.7ex\hbox{\gtsima}}
\def\simlt{\lower.7ex\hbox{\ltsima}}
\def\simpr{\lower.7ex\hbox{\prosima}}

\newcommand{\avg}[1]{\langle #1 \rangle}


%% file: citation_fix.tex
\usepackage{etoolbox}
\makeatletter
\patchcmd{\NAT@citex}
  {\@citea\NAT@hyper@{\NAT@nmfmt{\NAT@nm}\NAT@date}}
  {\@citea\NAT@nmfmt{\NAT@nm}\NAT@hyper@{\NAT@date}}
  {}
  {}
\patchcmd{\NAT@citex}
  {\@citea\NAT@hyper@{%
     \NAT@nmfmt{\NAT@nm}%
     \hyper@natlinkbreak{\NAT@aysep\NAT@spacechar}{\@citeb\@extra@b@citeb}%
     \NAT@date}}
  {\@citea\NAT@nmfmt{\NAT@nm}%
   \NAT@aysep\NAT@spacechar%
   \NAT@hyper@{\NAT@date}}
  {}
  {}
\patchcmd{\NAT@citex}
  {\@citea\NAT@hyper@{%
     \NAT@nmfmt{\NAT@nm}%
     \hyper@natlinkbreak{\NAT@spacechar\NAT@@open\if*#1*\else#1\NAT@spacechar\fi}%
       {\@citeb\@extra@b@citeb}%
     \NAT@date}}
  {\@citea\NAT@nmfmt{\NAT@nm}%
   \NAT@spacechar\NAT@@open\if*#1*\else#1\NAT@spacechar\fi%
   \NAT@hyper@{\NAT@date}}
  {}
  {}
\makeatother

%% file: macros.tex
\def\rsp{R_{\rm sp}}
\def\msp{M_{\rm sp}}

\def\astropy{\textsc{astropy}\xspace}